\documentclass[showpacs,twocolumn,superscriptaddress]{revtex4-1}

\usepackage[sc]{mathpazo}
\usepackage[T1]{fontenc}
\usepackage[hmarginratio=1:2,top=35mm,columnsep=1cm,hmargin=2cm]{geometry}
\usepackage{paralist}
\usepackage[titletoc,toc,title]{appendix}
\usepackage{amsmath,amssymb}
\usepackage{graphicx,subfigure,float,wrapfig}

\usepackage{titlesec}
\renewcommand\thesection{\Roman{section}}
\titleformat{\section}[block]{\large\scshape\centering}{\thesection.}{1em}{}

\usepackage{amsmath}
\usepackage{xcolor}
\usepackage{hyperref}
\hypersetup{
    colorlinks,
    linkcolor={red!50!black},
    citecolor={blue!50!black},
    urlcolor={blue!80!black}
}

\newcommand{\be}{\begin{equation}}
\newcommand{\ee}{\end{equation}}

\begin{document}

\title{Ballistic front dynamics after joining two semi-infinite quantum Ising chains}	

\author{%
	\textsc{Gabriele Perfetto$^{1,2},$ Andrea Gambassi$^{2,3}$}\\[2mm]
	$^1$DISAT, Politecnico di Torino, Corso Duca degli Abruzzi 24, 10129 Torino, Italy  \\
	$^2$SISSA - International School for Advanced Studies, via Bonomea 265, 34136 Trieste, Italy \\
    $^3$INFN, Sezione di Trieste, Trieste, Italy \\
	E-mail:  \href{Email:}{gperfetto@sissa.it, gambassi@sissa.it}}	

\begin{abstract}
We consider two semi-infinite quantum Ising chains initially at thermal equilibrium at two different temperatures and subsequently joined by an interaction between their end points. Transport properties such as the heat current are determined by the dynamics of the left- and right-moving fermionic quasi-particles which characterize the ensuing unitary dynamics. Within the so-called semi-classical space-time scaling limit we extend known results by determining the full space and time dependence of the density and current of energy and of fermionic quasi-particles.  Upon approaching the edge of the propagating front, these quantities as well as the two-point correlation function display qualitatively different behaviors depending on the transverse field of the chain being critical or not. While in the latter case corrections to the leading behavior are described, as expected, by the Airy kernel, in the former a novel scaling form emerges with universal features.
\end{abstract}

\maketitle


\section{Introduction} 
\label{sec:intro}

The non-equilibrium dynamics of isolated quantum statistical systems has recently captured a lot of attention from both the theoretical and the experimental point of view. In fact, significant experimental advances have made it possible to finely control trapped ultra-cold atomic gases (see, for instance, Refs.~\cite{1742-5468-2016-6-064001,eisert2015quantum,polkovnikov2011colloquium,altman2015non,calabrese2015non}). These systems are so weakly coupled with the surrounding environment that they allow the observation of a unitary non-equilibrium time evolution, with the consequent remarkable phenomena which cannot be observed in standard condensed matter systems due to decoherence and dissipative transport.

Within this context, one-dimensional systems reaching a non-equilibrium steady state and supporting a current, e.g., of energy, particles and charge, are now topical. In fact, on the one hand, they provide an approximation of actual three-dimensional systems with strong anisotropy and on the other, they display an anomalous heat conduction which violates Fourier's law because of the ballistic transport of energy. 
The role of spatial dimensionality in determining the features of the quantum dynamics of many-body systems out of equilibrium has been also experimentally demonstrated: for example, one-dimensional systems may relax towards a non-canonical distribution due to the possible presence of additional conservation laws which make them integrable compared to systems in higher dimensionality, see, e.g., Refs.~\cite{vidmar2016generalized,essler2016quench,ilievski2016quasilocal}. 
In this work, we investigate within this framework the non-equilibrium dynamics and transport properties of perhaps the best known integrable lattice model, i.e., the transverse field Ising chain (TFIC). In order to realize a non-equilibrium steady state (NESS) we adopt the protocol which involves two Hamiltonian reservoir, see, e.g., Refs.~\cite{bernard2012energy,spohn1977stationary}: the system consists of two adjacent TFICs, referred to as the left and the right chain, of finite length and initially disconnected and thermalized at two different temperatures $\beta_l^{-1}$ and $\beta_r^{-1}$, respectively. Apart from this different initial conditions, the two chains are otherwise identical. The initial probability distribution is encoded by the density operator 
\be
\rho_0 = e^{-\beta_r H_r} \otimes e^{-\beta_l H_l}/Z ,
\label{eq:intro-rho0}
\ee
where $H_r$ and $H_l$ are the Hamiltonians of the right and left chain, respectively, while $Z$ is a normalization constant. 
At time $t=0$ the two chains are instantaneously joined by switching on a local interaction between their closest end points, resulting in a unique homogeneous chain, twice as long as the two separate initial chains and crossed by a temperature gradient. The protocol just described in which the Hamiltonian governing the evolution of the system is subject to an abrupt change and afterwards the system evolves according to the unitary dynamics of the resulting Hamiltonian is generally referred to as a local quantum quench, see, e.g., Refs.~\cite{calabrese2011quantum,2012quantum,collura2013quench}.

The protocol described above has been extensively studied in various contexts: in particular, it has been shown that for critical one-dimensional quantum systems a NESS with a factorized density matrix emerges in terms of right- and 
left-moving excitations \cite{bernard2012energy,bernard2015non,doyon2012nonequilibrium,bernard2016conformal}. Similarly, the persistence of a non-equilibrium current at long times has been explained in terms of the 
rightward and leftwards 
ballistic propagation of the excitations of the initial state. These excitations enter into the adjacent chain from the contact point and  they establish a non-equilibrium state within a spatial region, the extent of which is determined by the velocity of the propagating front \cite{doyon2012nonequilibrium,bernard2016conformal}. The resulting energy current and the cumulant generating function of the energy transferred along the chain has been also determined in the long-time NESS \cite{bernard2012energy,bernard2015non}.

For models of free fermions, instead, the complete dynamics of the two-point correlation function and of some transport properties such as the energy density, the fermion concentration and the transverse magnetization has been obtained also in the transient regime preceding the NESS, both on a lattice with either an initial domain-wall state  \cite{eisler2013full,antal1999transport}  or a more general factorized Fermi sea state \cite{viti2015inhomogeneous} and on the continuum \cite{collura2014quantum,collura2014non}. For the specific case of the TFIC, the dynamics of the magnetization has been derived in Ref.~\cite{eisler2016universal} starting from a domain-wall initial state. In Ref.~\cite{de2013nonequilibrium}, instead, it has been shown that, starting from $\rho_0$ in Eq.~\eqref{eq:intro-rho0}, a NESS with a factorized density matrix develops also if the TFIC is not critical and the statistical properties of the stationary current (including the large-deviaton function) have bee
 n calculated analytically.

Here we consider the TFIC, not necessarily at its critical point, with an initial factorized thermal state described by $\rho_0$ in Eq.~\eqref{eq:intro-rho0} and we extend the aforementioned results by describing the complete space-time evolution of various relevant quantities such as the heat current, studying in detail how the eventual and known NESS is reached.  
Concerning the energy current, for example, we show that its non-analytic approach to the propagating front of the excitations depends qualitatively on whether the transverse field $h$ is critical ($h=h_c=1$) or not.
For the two-point correlation function we further investigate the behaviour upon approaching the edge of the front, showing that, due to the initial finite temperatures of the initially separate chains, these correlations acquire a correction compared to the case at zero temperature, known to be described by the Airy kernel \cite{tracy1994level}. For the energy current, as long as $h \neq h_c$, the edge behaviour turns out to be well described by the Airy kernel which determines a staircase structure of the profile beyond the semi-classical approach describing the dynamics far from the propagating front. 
For $h=h_c$, instead, such a profile changes qualitatively and in particular the staircase structure which characterizes the aforementioned Airy kernel is smoothed out and lost.   

The rest of the presentation is organized as follows: In Sec.~\ref{firstsection} and Appendix \ref{appendix} we briefly recall the exact solution of the TFIC,  following Ref.~\cite{de2013nonequilibrium}, in order to set the stage  for studying the non-equilibrium dynamics. In Sec.~\ref{secondsection}, we determine, within the space-time scaling limit, the time evolution of the relevant transport quantities, such as the energy current and the related energy density. The details of the corresponding calculations are reported in Appendix~\ref{appendix2}.
In Sec.~\ref{thirdsection}, we study the two-point correlation function and the energy current close to the edge of the propagating front, while the details of this analysis are collected in Appendix~\ref{appendix3}. Section~\ref{sec:comm} summarizes our results and presents our conclusions.  Part of this work is based on the unpublished results of Ref.~\cite{perf-thesis}.

\section{The Ising chain in a transverse field: exact solution} 
\label{firstsection}

In order to study the protocol discussed in the previous section, we assume that the two TFIC of length $N$ are originally disconnected, with the right one (denoted by the subscript $r$) occupying the lattice sites $\{1, 2,\ldots, N\}$ along a line, while the left one ($l$) the sites $\{-N+1,-N+2, \ldots, 0\}$. Accordingly, the corresponding Hamiltonians are, respectively,
\begin{subequations}
\label{chains}
\begin{align}
& H_r = -\frac{J}{2}\left[\sum_{q=1}^{N-1} \sigma_q^x \sigma_{q+1}^x  + h \sum_{q=1}^{N} \sigma_q^z \right], \\
& H_l = -\frac{J}{2}\left[\sum_{q=1}^{N-1} \sigma_{-q}^x \sigma_{-q+1}^x + h \sum_{q=0}^{N-1} \sigma_{-q}^z \right],
\end{align} 
\end{subequations}
where $\sigma_q^{x,y,z}$ are the usual Pauli matrices, $J$ is the coupling strength and $h$ the transverse field.  Open boundary conditions are assumed for both chains. The pre-quench Hamiltonian $H_0=H_l +H_r$ consists of the two disconnected and independent chains.

It is well known (see, e.g., Ref.~\cite{sachdev2007quantum}) that the first step in order to diagonalize this model consists of the Jordan-Wigner transformation:
\be
c_q = \left( e^{i \pi \sum_{l=1}^{q-1} {\sigma_l^+ \sigma_l^{-}}}\right)  \sigma_q^+ = \left(\prod_{l=1}^{q-1} \sigma_l^z\right) \sigma_q^+,
\label{JordanWigner}
\ee   
where we have introduced the usual spin raising and lowering operators $\sigma_q^\pm = (\sigma_q^x \pm i \sigma_q^y)/2$. In terms of these new fermionic operators $c_q$, with $\{c_q, c_{q'}^\dagger \} = \delta_{q,q'}$, the Hamiltonians $H_{r,l}$ acquire the bilinear form
\begin{equation}
H_r =  -\frac{J}{2} \sum_{q=1}^{N-1} \left(c_q^\dagger c_{q+1}^\dagger + c_q^\dagger c_{q+1} + h.c. \right)   +Jh\sum_{q=1}^{N} c_q^\dagger c_{q} ,
\label{Shultz}
\end{equation}
where $h.c.$ denotes the Hermitian conjugate of the preceding expression and an analogous form holds for $H_l$.  As discussed in Ref.~\cite{lieb2004two} and detailed in Appendix \ref{appendix}, Hamiltonians of this type can be diagonalized  via a Bogoliubov transformation which suitably introduces two fields $\phi_{r,l}(k)$ as
\begin{equation}
\phi_r (k) = \sum_{q=1}^{N} \left[ \omega_r^q (k) c_q + \xi_r^q(k) c_q^\dagger \right],
\label{phioperator}
\end{equation}
with an analogous expression for $\phi_l(k)$, but with coefficients $\{  \omega_l^q (k),\,  \xi_l^q(k)\}$ and $q=-N+1,\ldots, 0$. In terms of these fields one finds
\begin{equation}
H_{r,l} =  \sum_k \varepsilon(k) \phi_{r,l}^\dagger (k) \phi_{r,l} (k),
\label{diagonal}
\end{equation}
where the single-particle energy spectrum is given by 
\be
\varepsilon (k) = J\sqrt{h^2 -2h \cos k +1}. \label{spectrum}
\ee
Due to the finite length $N$ of both chains, the set of allowed values of $k$ in the sum of Eq.~\eqref{diagonal} is discrete and, as a consequence of the open boundary conditions, determined by the implicit condition 
\be
k_n = \frac{n \pi}{N+1} + \frac{f(k_n)}{N+1}, \quad \mbox{with} \quad n=0,1, ... N, 
\label{kvalues}
\ee
where
\be
f(k) \equiv \arctan\left(\frac{\sin k}{\cos k -h}\right). \label{angle}
\ee
In the thermodynamic limit $N \rightarrow \infty$, both chains become semi-infinite, either to their right or to their left. Correspondingly, the set of allowed values $k_n$ becomes continuous within the interval $[0,\pi]$ and, upon redefining ${\Phi_{r,l} (k)= \lim_ {N \rightarrow \infty} (N/\pi)^{1/2} \phi_{r,l} (k)}$, the Hamiltonians take the diagonal form
\begin{equation}
H_{r,l} = \int_0^\pi dk \; \varepsilon (k) \Phi_{r,l}^\dagger (k) \Phi_{r,l} (k),
\label{HRcont}
\end{equation}
where the single-particle energy spectrum $\varepsilon(k)$ is the same as in Eq.~\eqref{spectrum}.

At time $t=0$ the two chains are instantaneously joined in order to form a unique chain with Hamiltonian:
\begin{equation}
H= H_0 + \delta H = H_0 -\frac{J}{2} \sigma_0^x \sigma_1^x
\label{postquenchH-0}
\end{equation}
where $\delta H$ represents the energy cost of connecting the two chains through their closest end points at $q=0$ and $q=1$. Note that this operator is local, as it has support only across the connection between these two points. After the quench, the chain becomes translationally invariant in the thermodynamic limit and thus $[H,P_{tr}]=0$, where $P_{tr}$ is the translation operator along the chain defined by the action:
\begin{equation}
\sigma_{q-1}^\alpha = P_{tr}^\dagger \sigma_q^\alpha P_{tr}, \quad \mbox{with} \quad \alpha =x,y,z.
\end{equation}
Since $H$ is also invariant under the spatial inversion $P$, i.e., $[H,P]=0$, one realizes that for each value of the wavevector $k$ a two-fold degeneracy of the energy spectrum arises. Accordingly, one can introduce two fermionic operators $\Psi_{R,L}(k)$ (see Appendix \ref{appendix}) which are obtained via a suitable linear combinations of the pre-quench operators 
$\phi_r(k)$ and $\phi_l(k)$ (equivalently, $\Phi_r (k)$ and $\Phi_l (k)$) and which acquire opposite phases under the action of the translation operator:
\begin{equation}
P_{tr}^\dagger \Psi_{R,L} (k) P_{tr} = e^{\mp ik} \Psi_{R,L} (k);
\label{transinvariance}
\end{equation}
in terms of these $\Psi_{R,L}$, the post-quench Hamiltonian becomes: 
\begin{align}
H &= \int_0^\pi dk \; \, \varepsilon (k) \left[ \Psi_R^\dagger (k) \Psi_R (k) + \Psi_L^\dagger (k) \Psi_L (k) \right] \nonumber  \\
  &\equiv H_R + H_L .  \label{postquenchhamiltonian}
\end{align}

\section{Dynamics in the semi-classical limit} 
\label{secondsection}

As described in Sec.~\ref{sec:intro}, the NESS is obtained by joining at time $t=0$ the two chains, 
which are initially independently thermalized at two generically different temperatures and therefore have the density matrix $\rho_0$ reported in Eq.~\eqref{eq:intro-rho0}. In order to be able to access the stationary state, both the time $t$ and the system size $N$ must be large, the latter being always larger than the maximal distance $v_{max}t$ travelled by the fermionic excitations at time $t$, where $v_{max}$ is their maximal velocity (this quantity will be discussed further below in Eq.~\eqref{characteristicvel}).  
In this case, the steady state density matrix  $\rho_{stat}$ is formally defined by requiring that \cite{castro2014thermodynamic,bernard2015non,bernard2013time} 
\be
\langle O \rangle_{stat} \equiv \lim_{t \rightarrow \infty} \lim_{{\rm N} \rightarrow \infty} {\rm Tr}[O \rho(t)] = {\rm Tr}[O \rho_{stat}],
\label{eq:stat-av}  
\ee
for any generic local observable $O$, as discussed further below. Accordingly, $\rho_{stat}$ can be formally expressed as 
\be
\rho_{stat} = S \rho_0 S^\dagger, 
\label{stationary}  
\ee
where the operator
\be
S = \lim_{t \rightarrow \infty} \lim_{{\rm N} \rightarrow \infty} e^{-iHt} e^{iH_0t}  
\ee
evolves states to time $t=-\infty$ according to the dynamics of $H_0$ and then brings them back to $t=0$ with the dynamics prescribed by $H$. 
In order to observe the stationary behaviour, measurements have to be performed within the spatial region which has already reached a stationary state, the typical extension of which is given by $v_{max} t$, since excitations propagate ballistically, as we shall see below. 
As a consequence, the spatial support of the observable $O$ should include points at a maximal distance $\ell$ from the junction between the two chains which is much smaller than the distance $v_{max}t$ within which the steady state is established at time $t$, i.e.,  $\ell \ll vt \ll N$.
Under these conditions, Eq.~\eqref{stationary} defines the steady state $\rho_{stat}$ which describes the steady average of any operator $O$ with a finite support.

The protocol described above for realising a stationary state after joining two thermalised chains which act as asymptotic thermal baths, i.e.,  as Hamiltonian reservoirs, is usually referred to as partitioning protocol and it has been extensively studied \cite{bernard2016conformal,bernard2012energy,doyon2015non}: in particular, 
for this type of quench, it is well-known that the stationary density matrix $\rho_{stat}$  eventually takes the form
\begin{equation}
\rho_{stat} = e^{-\beta_r H_L} \otimes e^{-\beta_l H_R}/Z,
\label{noneqViti}
\end{equation} 
with $H_{L,R}$ given in Eq.~\eqref{postquenchhamiltonian}. Essentially, this tells us that the right- (viz.~left-)moving excitations of the Hamiltonian are characterised by the initial temperature of the left (viz.~right) chain.

\subsection{Energy current and density}
\label{sec:JEandu}

In the present work we will mainly focus on the dynamics of the energy current which emerges after joining the two chains.  In particular, in order to define the energy current flowing from the left to the right of the contact point $x=0$ of the two chains, it is natural \cite{bernard2016conformal} to consider the time variation of the energy $H_r$ of the right chain
\be 
j_E^r(0,t) = \frac{d H_r}{dt};
\ee
however, one can define also the energy current as the opposite of the energy variation of the left chain, i.e., as $j_E^l(0,t)= -d H_l / dt$. Here we consider the symmetric combination of these two equivalent contributions, that as well quantifies the total energy transferred from the left to the right chain across their junction link at time $t$
\be
j_E(0,t)= \frac{j_E^r(0,t) + j_E^l(0,t)}{2} = 
\frac{d E_0(t)}{dt},   
\ee
where we have introduced the energy difference  $E_0(t) \equiv (H_r-H_l)/2$ between the left and right chain. This definition can be naturally extended to a generic point $x$ of the chain with $j_E (x,t) = d E_x(t)/dt$ being the energy current across the link between sites at position $x$ and $x+1$ along the chain. Specifically, $j_E$ takes the form 
\be
j_E(x,t) = \frac{ihJ^2}{2} e^{iHt} \left(c_{x+1}^\dagger c_x - c_x^\dagger c_{x+1} \right) e^{-iHt}. \label{localcurrent}
\ee
Accordingly, the average energy current $J_E(x,t)$ at time $t$ and point $x$ along the chain is given by
\be
J_E (x,t) = \mbox{Tr}[j_E(x,t) \rho_0]. 
\label{trace}
\ee
The stationary and space-independent average value $J_E^{NESS}$ of the current operator $J_E(x,t)$ in the NESS specified by $\rho_{stat}$ in Eq.~\eqref{noneqViti} was calculated in Ref.~\cite{de2013nonequilibrium} according to the prescription of Eq.~\eqref{eq:stat-av} and it turns out to depend on the initial inverse temperatures 
as \cite{karrasch2013nonequilibrium}, 
\begin{equation}
J_E^{NESS} = g(\beta_l)-g(\beta_r), 
\label{vitiresult}
\end{equation}
where the function $g$ will be discussed further below, after Eq.~\eqref{eq:JENESS-here}. 
However, the approach of $J_E (x,t)$ to $J_E^{NESS}$ was not previously investigated and here we fill in this gap. 
Equation~\eqref{vitiresult} can be specialised to the case of chains with a critical field $h=h_c=1$ in Eq.~\eqref{chains}
and, in the scaling limit $\beta_{r,l}J \gg 1$, it turns out to agree with the general prediction $J_E^{NESS} = \pi(\beta_l^{-2} - \beta_r^{-2})/24$ of conformal field theory \cite{bernard2012energy,bernard2015non,biella2016energy,castro2016emergent,de2014energy}.

In order to determine the dynamics of the current operator in Eq.~\eqref{localcurrent} and other similar observables, the approach described above --- based on Eq.~\eqref{noneqViti} --- is not viable, as it provides information only on the NESS. Accordingly, we directly calculate the average of space-time dependent observables from the initial density operator: in particular, the energy current $J_E(x,t)$ is determined from the trace in Eq.~\eqref{trace}. In order to do this, one first expresses the operator $j_E$ in Eq.~\eqref{localcurrent} in terms of right- and left-moving fermions $\Psi_{R,L}$ appearing in Eq.~\eqref{postquenchhamiltonian}, by inverting the transformations reported in Sec.~\ref{firstsection} which relate them to the original fermionic operators $c_x$, as detailed in Appendix~\ref{appendix}. This is done via the following Bogoliubov transformation: 
\begin{equation}
c_x = \int_{-\pi}^\pi dk \left[ \Psi_{R} (k) \left({\omega_R^x (k)}\right)^\ast+ \Psi_R^{\dagger} (k) \xi_R^x (k)  \right], \label{fourier}    
\end{equation}
where ${\omega_R^x (k)}$ and $\xi_R^x (k)$ are given in Eq.~\eqref{opposite}. 
Remembering that the dynamics of $\Psi_R(k)$ under
the post-quench Hamiltonian $H$ in Eq.~\eqref{postquenchhamiltonian} is trivial, i.e., $e^{-iHt} \Psi_R (k) e^{iHt}=e^{i\varepsilon(k)t} \Psi_R (k)$, the time evolution of $j_E$ in Eq.~(\ref{localcurrent}) can be explicitly determined (see Appendix \ref{appendix2}). The remaining average over the initial density matrix turns out to be
\be
J_E (x,t) =  \int_{-\pi}^{\pi} \frac{dk dk'}{2 \pi}\frac{ihJ^2}{2}\left(e^{-ik}-e^{ik'}\right)I(k,k')e^{i \varphi_{x,t}(k,k')} \label{operator2}
\ee
where 
\be
\varphi_{x,t}(k,k')=[\varepsilon(k)-\varepsilon(k')]t+x(k'-k),
\ee
with $\varepsilon(k)$ defined in Eq.~\eqref{spectrum}, while $I(k,k')=\mbox{Tr}\left[\rho_0 \Psi_R^{\dagger} (k) \Psi_R(k') \right]$ (see Eq.~\eqref{integralapp}) encodes the information about the initial state and is given by Eq.~\eqref{integraldef}. 
In particular, we consider the so-called space-time scaling limit, also referred to as semi-classical or hydrodynamic approach \cite{viti2015inhomogeneous,collura2014quantum,collura2014non,eisler2016universal,bertini2016transport,castro2016emergent}, in which both the time $t$ and the coordinate $x$ are assumed to be much larger than the corresponding microscopic scales, respectively set by $J^{-1}$ --- the inverse of the energy of a single link of the chain --- and $v_{max}J^{-1}$ --- the typical velocity of the excitations, introduced further below in Eq.~\eqref{characteristicvel} --- but such that the ratio $x/t$ takes arbitrary finite values. In this limit, 
$J_E(x,t)$ in Eq.~\eqref{operator2} is determined by the values of $k$ and $k'$ within the integration domains at which the phase $\varphi_{x,t}(k,k')$ in the exponential is stationary and by the possible singularities of $I(k,k')$. 
Since $\varphi_{x,t}(k,k')$ turns out to be stationary for $k= k'$, the integral is then determined by the behaviour of $I(k,k')$ for $k\simeq k'$. Accordingly, it is convenient to introduce the variables $Q= k-k'$ and $K= (k+k')/2$ and to consider the integrand in Eq.~\eqref{operator2} for $Q\simeq0$ following the procedure highlighted in Ref.~\cite{viti2015inhomogeneous}. Expanding the phase $\varphi_{x,t}$ up to first order in $Q$ and $K$ one eventually gets:
\be
\begin{split}
&J_E (x,t) =  \int_{-\pi}^{\pi} \frac{dk}{2 \pi} \varepsilon(k) v_g(k)   \\
&\ \ \times \left[ f_{\beta_l}(k)\Theta(v_g(k)t-x)
          -    f_{\beta_r}(k)\Theta(v_g(k)t + x)\right],   
\label{finalresult}
\end{split}
\ee
where $v_g(k)=d\varepsilon(k)/dk$ is the group velocity of the relevant excitations, 
with $\varepsilon(k)$ in Eq.~\eqref{spectrum}, and
$f_{\beta}(k)=1/[1+e^{\beta \varepsilon(k)}]$ is the usual Fermi-Dirac distribution at inverse temperature $\beta$ which encodes the distributions of quasi-particles in the chains before they are joined. In Eq.~\eqref{finalresult}, $\Theta(v)$ indicates the step function with $\Theta(v\ge 0)=1$ and $\Theta(v<0)=0$.
This equation gives the exact profile of the average energy current at a certain time $t$ and point $x$ along the chain within the space-time scaling regime. Exploiting the continuity equation 
\be
\frac{\partial}{ \partial t} u(x,t) =- \frac{\partial}{\partial x} J_E(x,t),
\label{eq:continuity}
\ee 
expected to hold for the heat current, we can calculate the energy density $u(x,t)$ along the chain, with a temporal integration of the r.h.s.~of this equation. The initial condition $u(x,t=0)$ required in the integration derives from the fact that at $t=0$ the quasi-particles with energy $\varepsilon(k)$ are distributed according to the distribution $f_{\beta_l}(k)$ on the left chain $x \leq 0$ and to  $f_{\beta_r}(k)$ on the right one $x > 0$, i.e.,  
\begin{equation}
u(x,0) = \int_{-\pi}^{\pi} \frac{dk}{2 \pi} \varepsilon(k) [ f_{\beta_l}(k) \Theta(-x) + f_{\beta_r}(k) \Theta(x)]. \label{initialcondition}
\end{equation}
After integration in time of Eq.~\eqref{eq:continuity} and the restriction of the domain of integration, one eventually finds
\begin{equation} 
\begin{split}
&u(x,t) =  \int_{0}^{\pi} \frac{dk}{2 \pi} \varepsilon(k) \big\{ f_{\beta_l}(k)+f_{\beta_r}(k) \\
&+\left[f_{\beta_l}(k)-f_{\beta_r}(k)\right]\left[\Theta(v_g(k)t-x) -\Theta(v_g(k)t+x)\right] \big\}.  
\label{finalresultenergy}
\end{split}
\end{equation}
Expressions similar to Eqs.~\eqref{finalresult} and \eqref{finalresultenergy} have been obtained with different approaches for other spin models in one spatial dimension: 
specifically, for the XX model (equivalent to a model of free fermions) it has been found \cite{antal1999transport} that the magnetization and magnetization current evolve in the space-time scaling limit according to a scaling function which, apart from constants and prefactors, is similar to Eq.~\eqref{arcos} and \eqref{semicircular}, respectively, discussed further below.  
For the TFIC in an initial thermal tensor state (of the form in Eq.~\eqref{eq:intro-rho0}) for the two halves of the chain, 
instead, 
the space-time dependence of the transverse magnetization $m(q,t) = \langle \sigma_q^z(t) \rangle$  has been numerically studied \cite{platini2006out} for $h=h_c=1$, with one vanishing and one infinite initial temperature. The ensuing wavefront exhibits a light cone analogous to the one analyzed in this work, while the interpolation between the asymptotic values of $m(q,t)$ for $|q|>v_{max} t$ occurs linearly as a function of $q/t$. 
%
%
\begin{figure}
\centering
\includegraphics[width=1.2\columnwidth]{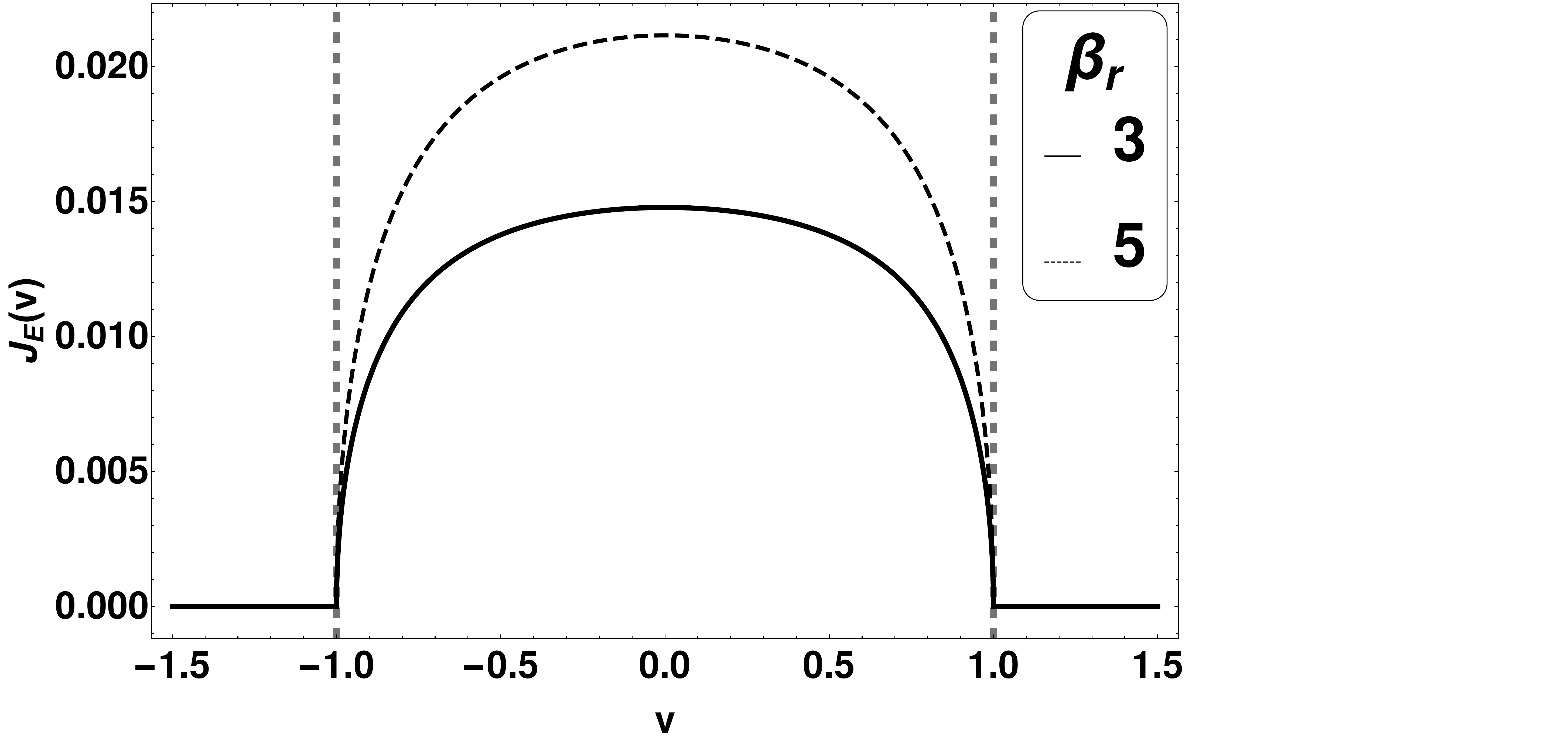}\\[-1mm]	
	(a)\\[2mm]
\includegraphics[width=1.28\columnwidth]{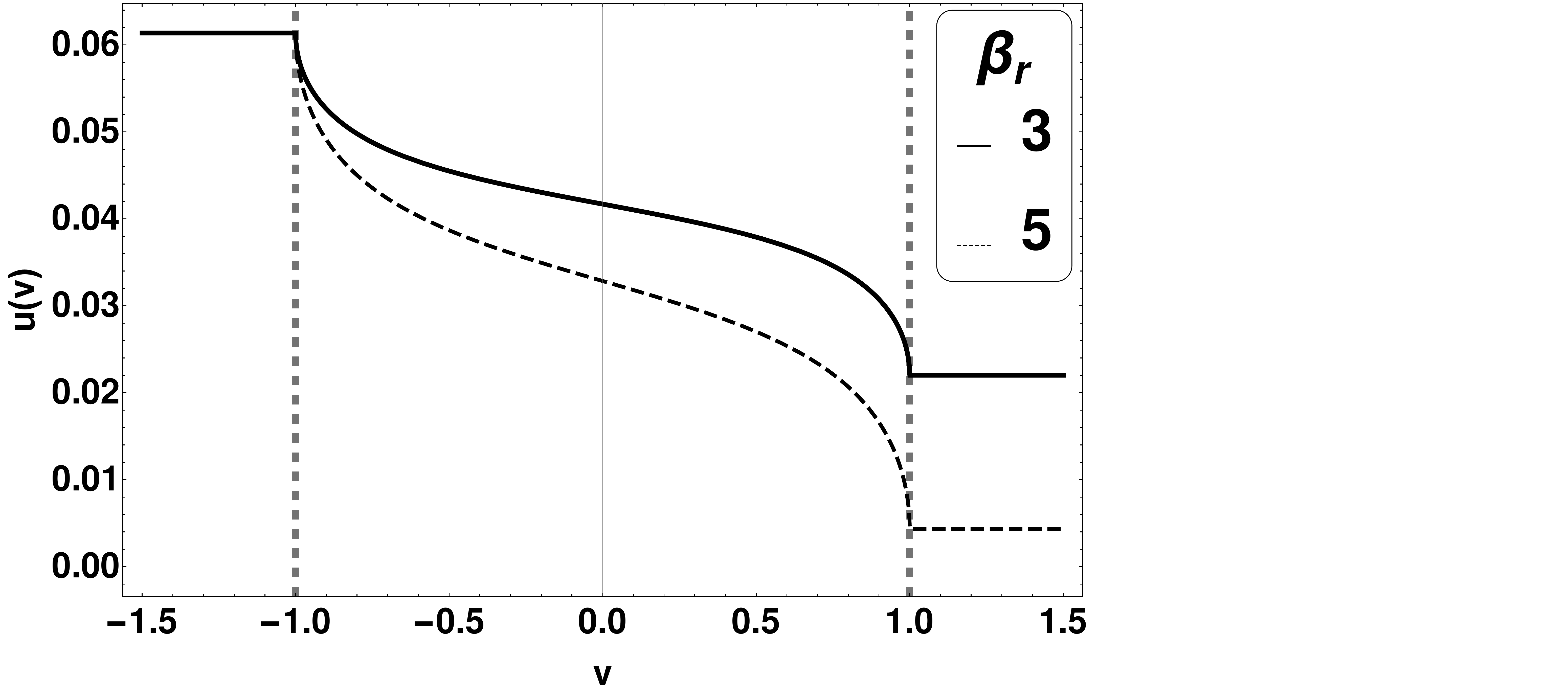}\\[-1mm]
(b)
\caption{Dependence of (a) the heat current $J_E$ and (b) the energy density $u$ at time $t$ and point $x$ along the Ising chain on the scaling variable $v\equiv x/t$, within the space-time scaling limit. The parameters of the chain are $h =1.6$ and $J=1$, with the left part $v<0$ initially thermalised at $\beta_l = 2$ and the right part $v>0$ at $\beta_r=3$ (thick solid line) and $5$ (dashed line).} 
\label{fig:scaling}
\end{figure}
%
%
Since the space and time dependence of the current in Eq.~\eqref{finalresult} is fully encoded by the two functions $\Theta$,  their arguments can be rescaled by a factor $t>0$ and therefore $J_E(x,t)$ turns out to be a function solely of the scaling variable $v \equiv x/t$, with $v \rightarrow \pm \infty$ corresponding to both short times and large distances and $v \rightarrow 0$ to long times or short distances from the origin. This dependence is reported in Fig.~\ref{fig:scaling} for both the heat current $J_E(x,t) \equiv J_E(v)$ and the associated energy density $u(x,t) \equiv u(v)$ for two different values of the inverse temperature $\beta_r$ with fixed $\beta_l=2$  (all measured in units of $J^{-1}$). 
The profile of the heat current turns out to be an even function of $v$, i.e., $J_E(v)=J_E(-v)$, as it is clearly shown by the figure and by a careful inspection of Eq.~\eqref{finalresult}, see the discussion after Eq.~\eqref{finalresultapp}. 
In addition, Fig.~\ref{fig:scaling} shows that, because of causality and the finite maximum value $v_{max}$ of the propagation velocity $v_g(k)$ of the excitations, there is always a region in space within which the initial state is not perturbed and, correspondingly, $J_E$ vanishes. These regions act as unperturbed "thermal reservoirs" for the central part of the chain and, with the dispersion relation in Eq.~\eqref{spectrum}, one finds (see also Ref.~\cite{2012quantum}) 
\be
v_{max} = \max_{k\in [-\pi, \pi]}\vert{v_g (k)}\vert = J \min(h,1). 
\label{characteristicvel}
\ee       
Accordingly, the "front" of the perturbation due to joining the two chains propagates with velocity $v_{max}$ and the central perturbed region expands with velocity $2 v_{max}$. Note that $v_{max}$ is at most $J$ and with $J=1$ and $h>1$ it takes the value $v_{max}=1$, as it is clearly shown by the plots in Fig.~\ref{fig:scaling}. 

Close to the joining point  $x=0$ of the two initial chains, where the quench has been performed, the profile is, instead, approximately flat with a value $J_E(v=0)$ corresponding to the one $J^{NESS}_E(\beta_l,\beta_r)$ eventually attained in the stationary state reached uniformly over the entire chain; indeed, the value $J_E(v=0)$ resulting from Eq.~\eqref{finalresult} coincides with $J^{NESS}_E(\beta_l,\beta_r)$ reported in Eq.~\eqref{vitiresult} and determined in Ref.~\cite{de2013nonequilibrium}.
Equation \eqref{finalresult} extends this known result on the steady-state value of the current as to describe its transient behaviour in both space and time, i.e., it predicts how the far region of the thermal reservoirs and the steady state value are asymptotically approached by the dynamics.

The physical interpretation of Eq.~\eqref{finalresult} is straightforward in terms of the propagation of quasi-particles with momentum $k$ and velocity $\pm v_g(k)$ which are produced in the initial state with a statistics $f_{\beta_r}(k)$ and $f_{\beta_l}(k)$ for $x>0$ and $x<0$, respectively, and which contribute with $\varepsilon(k) v_g(k) d k$ to the flow of energy.  In particular, this interpretation has fist been proposed for the TFIC in Ref.~\cite{sachdev1997low} and then used in Ref.~\cite{eisler2016universal} for the same model and in Ref.~\cite{antal2008logarithmic} for the XXZ chain. In fact, since the post-quench Hamiltonian $H$ in Eq.~\eqref{postquenchhamiltonian} is diagonal in terms of the operator $\Psi_{R}(k)$, the states $ \vert k \rangle_{R} = \Psi_{R}^\dagger (k) \vert 0 \rangle$ of the Fock space 
(where $\vert 0 \rangle$ indicates the ground state of the chain)
have an infinite lifetime and therefore propagate freely, with no scattering. Based on this picture, Eq.~\eqref{finalresult} (as well as all the analogous equations which are presented further below) could have been derived without the explicit calculations reported above. In fact,
%
\begin{figure}
\begin{center}
\includegraphics[width=0.8\columnwidth]{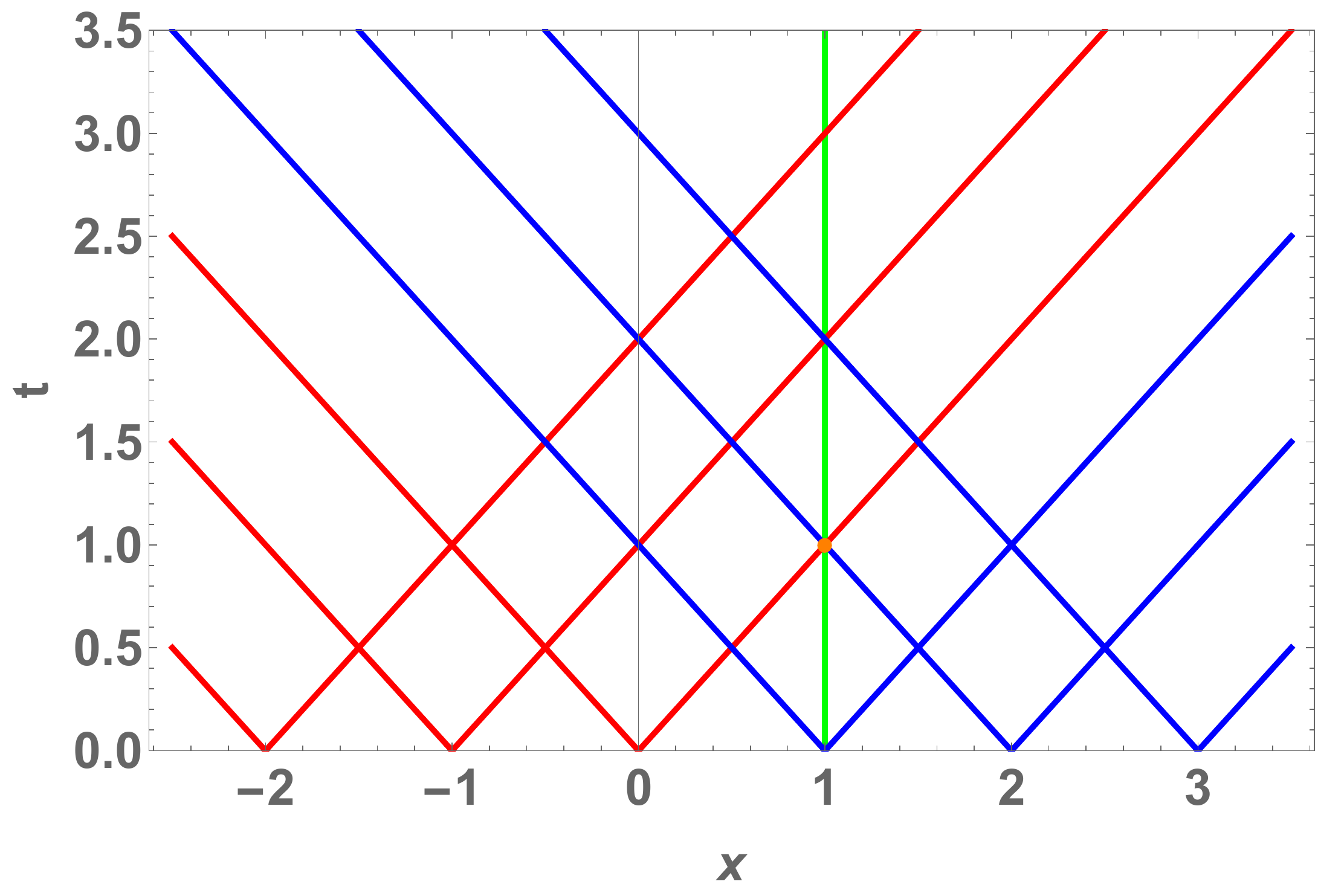}	
\end{center}
\caption{Space-time diagram in which the coordinate $x$ along the chain and the time $t>0$ are reported on the horizontal and vertical axis, respectively. Each point of the chain at time $t=0$ acts as a source of quasi-particles with velocities $\pm v_g(k)$, with $k>0$, energy $\varepsilon(k)$ and statistics $f_{\beta_l}(k)$ for $x<0$ (red rays) and   $f_{\beta_r}(k)$ for $x>0$ (blue rays), respectively. This particles propagate ballistically for $t>0$ and generate an energy current, as discussed in the main text. The green vertical line represents the "world line" of the fixed point $x=1$ in this space-time diagram.}
\label{fig:lightcone}
\end{figure}
consider the space-time diagram in Fig.~\ref{fig:lightcone}: the excitations with wavevector $k>0$ produced uniformly along the chain at time $t=0$ travel ballistically with velocity $\pm v_g(k)$ for $t>0$ and, in particular, those with statistics $f_{\beta_r}(k)$ (blue rays in Fig.~\ref{fig:lightcone}) [viz.~$f_{\beta_l}(k)$ (red rays)] originating from $x>0$ [viz.~$x<0$] also propagate into the complementary part of the chain. As a result, the flux of energy (i.e., the energy current) produced by each of these modes at a point with coordinate $x$ (e.g., $x=1$ in Fig.~\ref{fig:lightcone}, corresponding to the green vertical world line) vanishes for $| v_g(k) | t  < |x|$ because the flux of energy carried by the particles with wavevector $+k$ cancels out the one of particles with wavevector $-k$ moving in the opposite direction and having the same statistics. This cancellation no longer occurs for  $| v_g(k) | t  > |x|$ because, for $x>0$, the statistics of the excitations with veloc
 ity $-v_g(k)$ 
crossing the world line  of the point $x$ (in green in Fig.~\ref{fig:lightcone}) is given by $f_{\beta_r}(k)$ while that of the excitations with velocity $+v_g(k)$ by $f_{\beta_l}(k)$, as they were originally generated in the left part of the chain, see the sketch in Fig.~\ref{fig:lightcone}.  
As a consequence, for each value of $k\in [0,\pi]$, the contribution to the energy flux  is given by $\varepsilon(k) v_g(k) d k \times [f_{\beta_l}(k) - f_{\beta_r}(k)]\Theta(v_g(k)t-x)$ for $x>0$ and $\varepsilon(k) v_g(k) d k \times [f_{\beta_l}(k) - f_{\beta_r}(k)]\Theta(v_g(k)t+x)$ for $x<0$, which is equivalent to the integrand of Eq.~\eqref{finalresult}. 
Analogous interpretation can be given to the expression for the energy density $u(x,t)$ reported in Eq.~\eqref{finalresultenergy}, which can be actually derived without the analysis presented above in this and in the previous section.

The integral over $k$ in Eq.~\eqref{finalresult} can  be calculated in analytic form, as detailed in Appendix \ref{appendix2}, and therefore $J_E(v)$ can be written in the form
\begin{equation}
J_E(v)= \Theta (v_{max} - |v|) 
\left[{\cal J}_1(\beta_l,v)-{\cal J}_1(\beta_r,v)\right]
\label{rl}
\end{equation} 
where 
\be
\begin{split}
{\cal J}_1(\beta,v)= &\frac{1}{2 \pi \beta^2}\left\{ G_1(\beta [\varepsilon_>(v)-\varepsilon_<(v)])+\right. \\
                    & \qquad \left. -G_1(\beta [\varepsilon_>(v)+\varepsilon_<(v)])\right\},
\label{integration}
\end{split}
\ee
with 
\be
G_1(x)= - \mbox{Li}_2(-e^{-x})+x \log(1+e^{-x}),
\label{eq:def-G-mt}
\ee
$\varepsilon_>(v) = \sqrt{[J \mbox{max}(1,h)]^2- v^2}$, and $\varepsilon_<(v)$ has the same expression as $\varepsilon_>$ but with max replaced by min, such that (see Eq.~\eqref{characteristicvel}) $\varepsilon_<(v) = \sqrt{v_{max}^2-v^2}$.
Since $\varepsilon_\gtrless(v)$ and therefore ${\cal J}_E(v)$ depends only on $v^2$, the energy current in Eq.~\eqref{rl} is confirmed to be an even function of $v$, as anticipated above.

\begin{figure}
\begin{center}
\includegraphics[width=0.8\columnwidth]{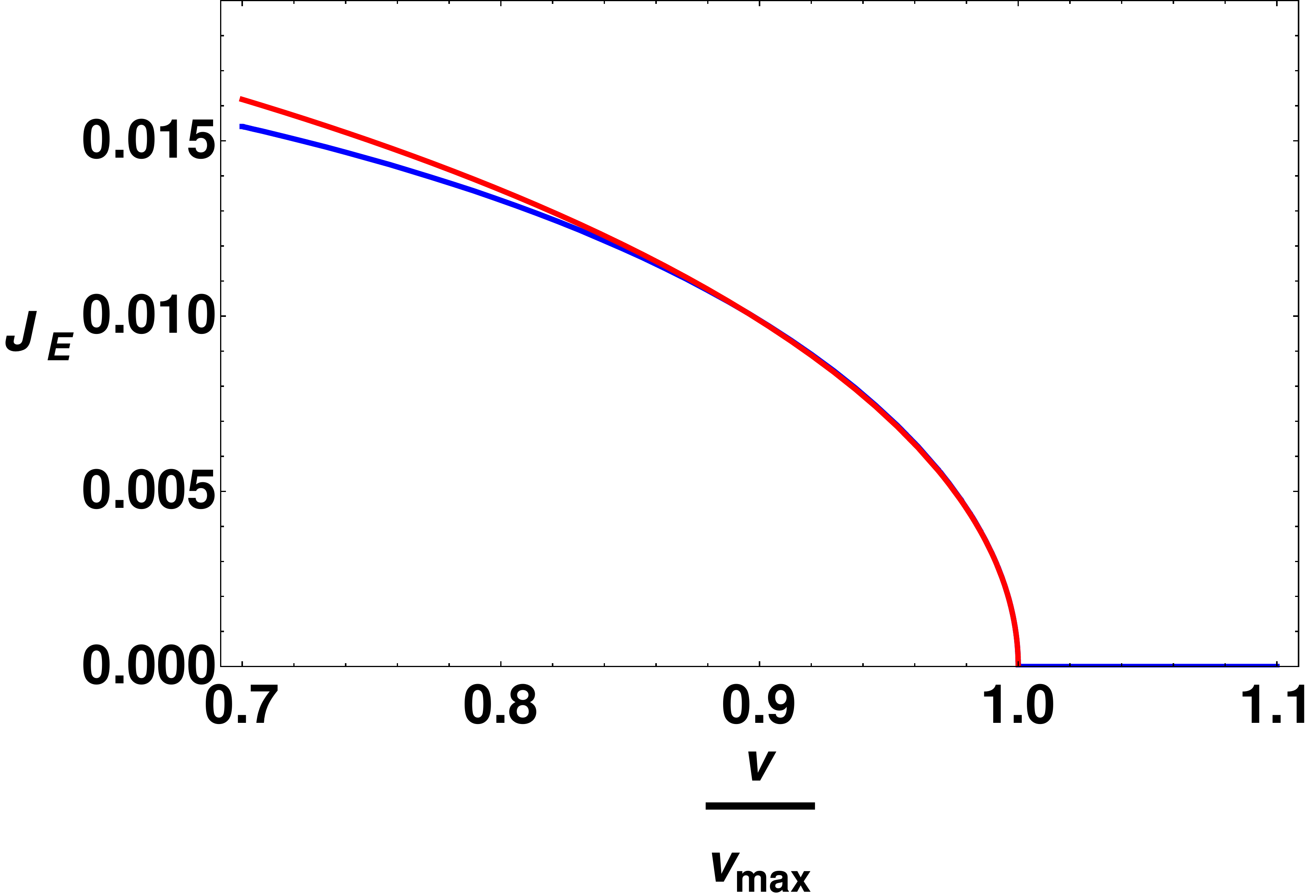}\\[-1mm]	
(a)\\[2mm]
\includegraphics[width=0.8\columnwidth]{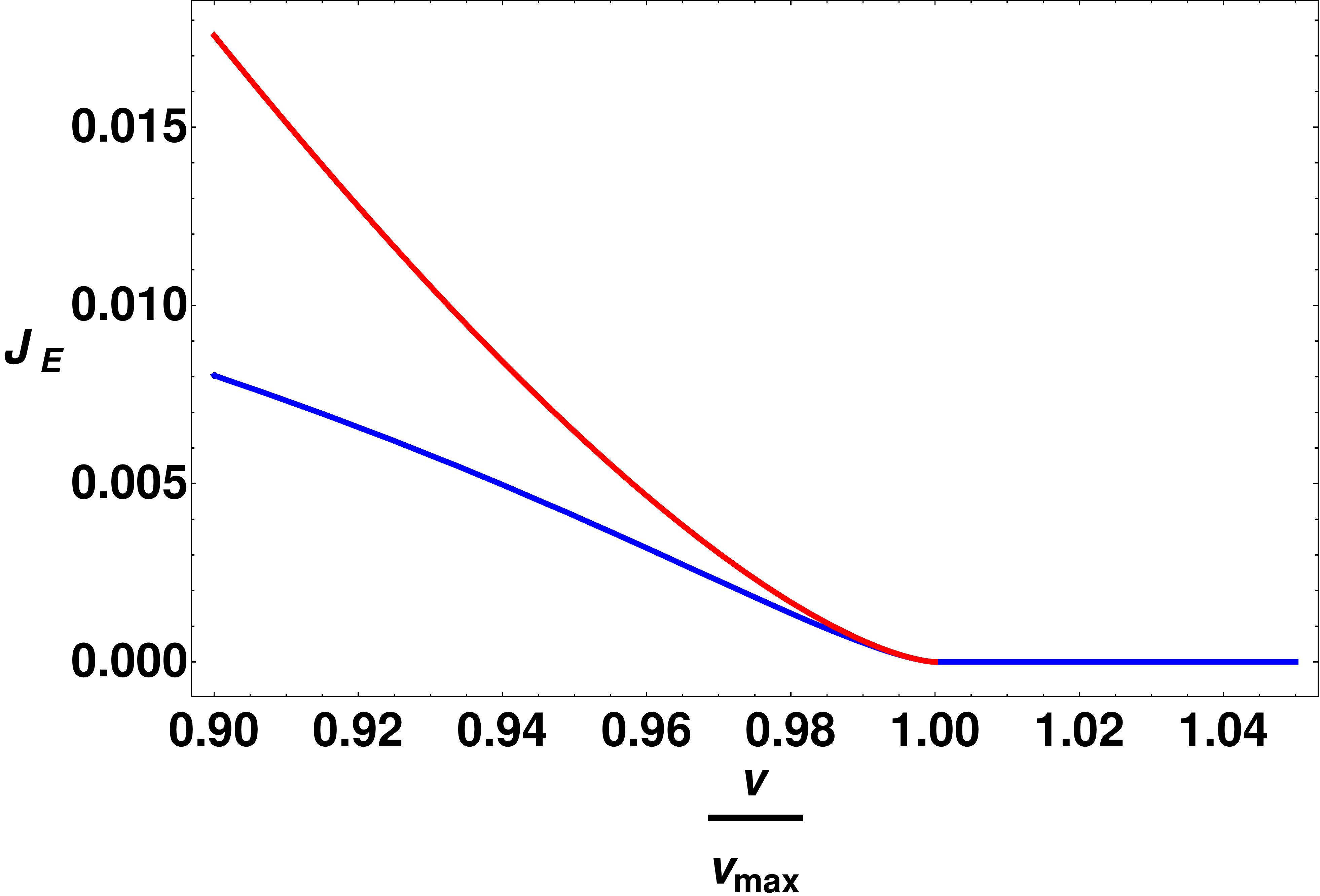}\\[-1mm]
(b)
\end{center}	
\caption{Energy current $J_E(v)$ as a function of the scaling variable $v/v_{max}$ for a TFIC with (a) $h=0.8$ and (b) $h=h_c=1$, prepared in an initial state with $\beta_r=4$ and $\beta_l=2$, where $J=1$ is assumed in both cases. The lower (blue) curves in both panels correspond to Eq.~\eqref{rl}, while the upper (red) curves to their approximation close to the edge $v \rightarrow v_{max}^- = Jh$ in Eqs.~\eqref{semicircular} and \eqref{eq:edge-crit} for panels (a) and (b), respectively.}
	\label{fig:edge1}
\end{figure}

Figure~\ref{fig:scaling} clearly shows that $J_E(v)$, upon approaching the values $\pm v_{max}$ of the variable $v$ which correspond to the edge of the propagating front, displays a non-analytic behaviour, which can be determined on the basis of Eqs.~\eqref{rl} and \eqref{integration}.  
In particular, for $v \rightarrow \pm v_{max}^\mp$ one finds, at the leading order, 
\be
J_E(v) = C_1 \left( v_{max}^2-v^2 \right)^{1/2} + {\cal O}((v_{max}-|v|)^{3/2}),  
\label{semicircular}
\ee
where $C_1$ is given in Eq.~\eqref{eq:genexprC} [see also Eq.~\eqref{C1constant}] and depends on $h$ and $\beta_{r,l}$.
Note that $J_E(v)$ vanishes at the edge according to a semi-circular law, as shown in Fig.~\ref{fig:edge1}(a), and consistently with what is observed in Ref.~\cite{antal1999transport} for the XX chain evolving from a domain-wall initial state and in Ref.~\cite{eisler2016universal} for the TFIC in an initial domain-wall state created by the action of a local Jordan-Wigner fermion operator.
However, when the transverse field $h$ of the Ising chain is poised at its critical value $h_c=1$, the constant $C_1$ in Eq.~\eqref{semicircular} vanishes and the approach of $J_E(v)$ to the edge turns out to change qualitatively (see the discussion after Eq.~\eqref{eq:genexprC}), with
\be
\begin{split}
J_E(v) =  & \frac{\beta_r-\beta_l}{3\pi} \left(v^2_{max}-v^2\right)^{3/2}
 + {\cal O}((v_{max}-|v|)^{5/2}). 
\end{split}
\label{eq:edge-crit}
\ee
This is clearly shown in Fig.~\ref{fig:edge1}(b), where we plot the behaviour of $J_E(v \to v_{max}^-)$ for the same conditions as in panel (a), but with $h=1$ and we compare it with the prediction in Eq.~\eqref{eq:edge-crit}.

As anticipated above, we note that the stationary value $J_E^{NESS}$ of the current  within the space-time scaling limit corresponds to $J_E(v=0)$: from Eqs.~\eqref{rl} and \eqref{integration} it takes the form
\be
\begin{split}
J_E^{NESS} =&  \frac{G_1(\beta_l J |h-1| ) - G_1(\beta_l J (h+1) )}{2\pi\beta^2_l} \\
&-  \frac{G_1(\beta_r J |h-1| ) - G_1(\beta_r J (h+1) )}{2\pi\beta^2_r},
\end{split}
\label{eq:JENESS-here}
\ee
which reproduces the known expression reported in Eq.~(24) of Ref.~\cite{ de2013nonequilibrium} (which assumes $h>1$), once one recognizes that $G_1(x)$ here equals $-j(x)$ therein. 

\begin{figure}
\begin{center}
\includegraphics[width=1\columnwidth]{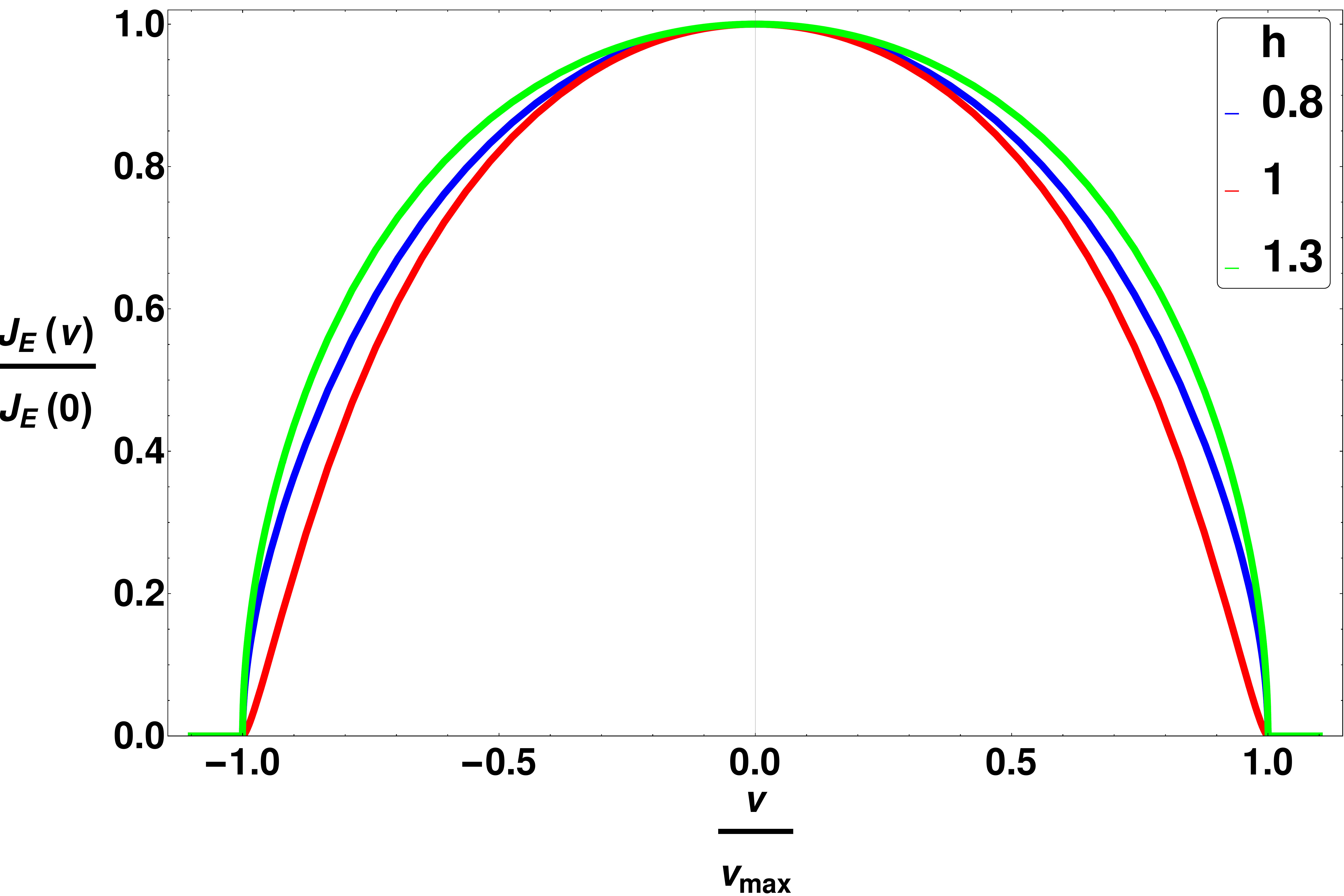}	
\end{center}
\caption{Dependence of $J_E(v)/J_E(0)$ on $v/v_{max}$ for $h=1.3$, $h=0.8$, and $h=h_c=1$, from top to bottom, for the TFIC with $J=1$, $\beta_l=1$, and $\beta_r=3.5$.}
\label{fig:color}
\end{figure}

In order to highlight the qualitative differences in the normalized profile $J_E(v)/J_E^{NESS}$ (with $J_E^{NESS} = J_E(0)$) as a function of $v/v_{max}$ upon varying $h$, we plot it in Fig.~\ref{fig:color} for three different values of the magnetic filed $h$. Within the central part of the interval, the stationary state has already been reached and in fact the curve approaches one. Near the edges, instead, the behaviour of the normalized current changes significantly at the critical point $h=h_c=1$, as the curve approaches these edges with a vanishing slope, differently from the non-critical case where the semicircular behaviour of 
Eq.~\eqref{semicircular} causes its divergence.    
%
%
%
%
%
%

\begin{figure}
\begin{center}
\includegraphics[width=0.8\columnwidth]{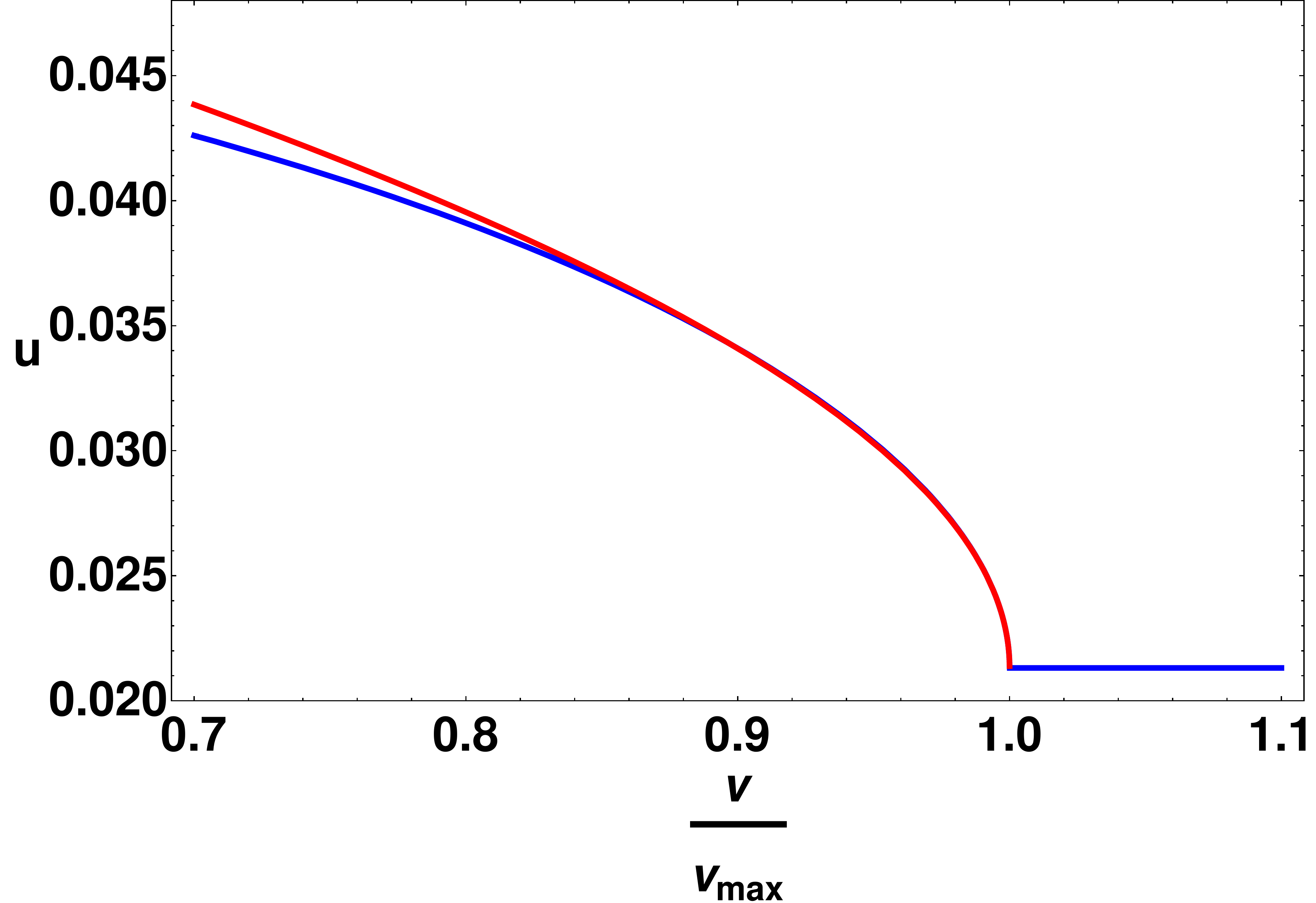}\\[-1mm]	
(a)\\[2mm]
\includegraphics[width=0.8\columnwidth]{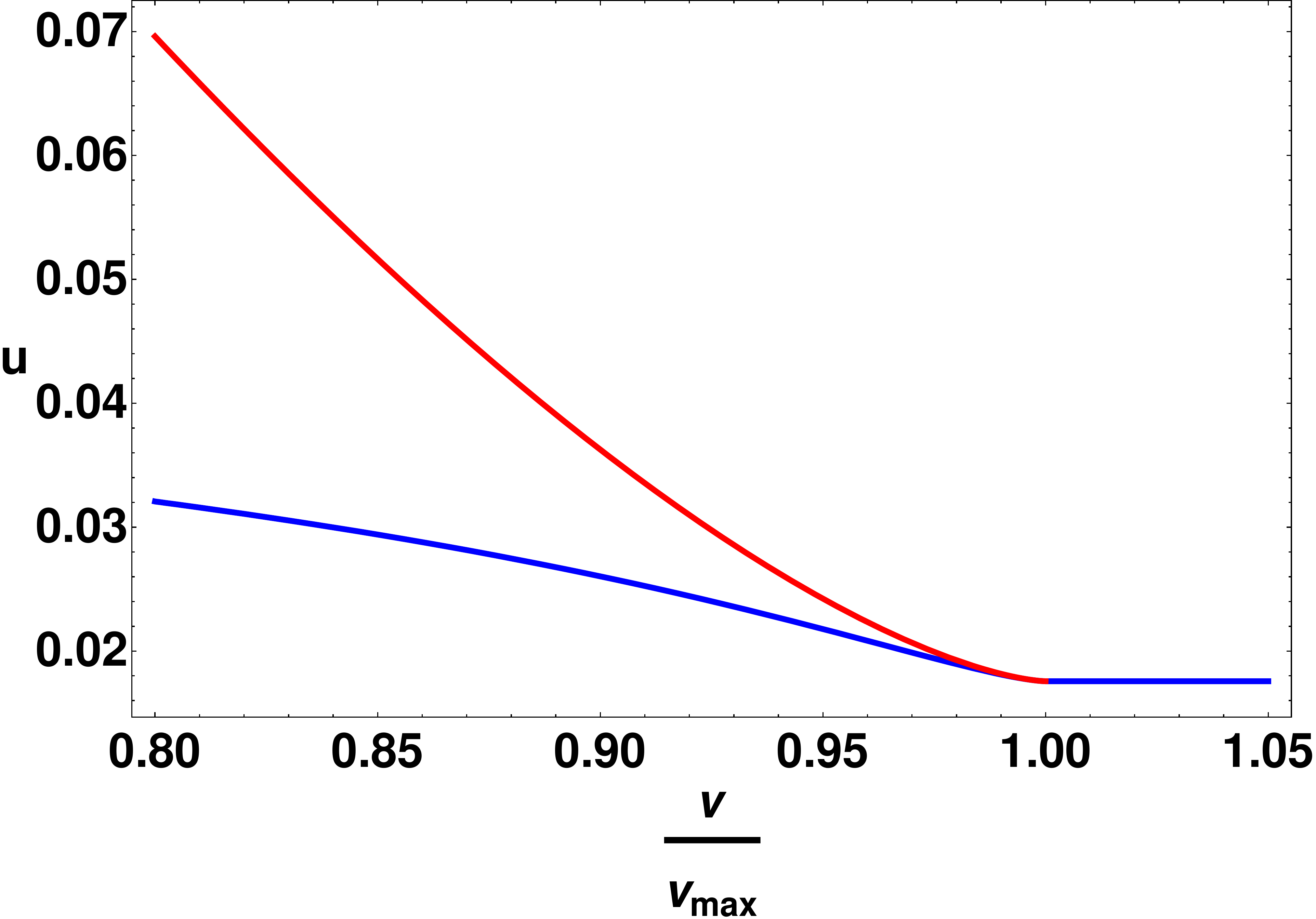}\\
(b)
\end{center}	
\caption{Energy density $u(v)$ as a function of the scaling variable $v/v_{max}$ for the same values of the parameters as those in the corresponding panels of Fig.~\ref{fig:edge1}. In both panels, the lower (blue) curves correspond to Eq.~\eqref{finalresultenergy}, while the upper (red) curves to their approximation close to the edge $v \rightarrow v_{max}^- = Jh$ in Eqs.~\eqref{arcos} and \eqref{arccos-crit} for panels (a) and (b), respectively.}
\label{fig:edge2}
\end{figure}

Exploiting the continuity equation  \eqref{eq:continuity}, rewritten in terms of the scaling variables
\be
\frac{d}{dv} u(v) = \frac{1}{v}\frac{d}{dv} J_E(v),
\label{eq:cont-dim}
\ee
Eq.~(\ref{semicircular}) can be used in order to derive the edge behaviour of $u(v)$; for $h\neq 1$, it turns out to be
\be
\begin{split}
u(v\to \pm v_{max}^\mp)=u(\pm\infty)\pm & C_1 \; \mbox{arccos}\left(\frac{|v|}{v_{max}}\right)\\ 
&+ {\cal O}((v_{max}-|v|)^{3/2}),
\label{arcos}
\end{split}
\ee
where $u(\pm\infty)$ is the value of the spatially constant initial energy density of the chain for $x>0$ and $x<0$, 
determined by the initial temperatures $\beta_r^{-1}$ and $\beta_l^{-1}$, respectively [the explicit expression of $u(x,t=0)$ is reported in Eq.~\eqref{initialcondition}]. 
Equation~\eqref{arcos} is compared in Fig.~\ref{fig:edge2}(a) with the actual energy density profile which can be obtained from Eq.~\eqref{finalresultenergy} and it displays the same qualitative features as the recent result of Ref.~\cite{eisler2016universal} concerning the magnetization $\langle \sigma_n^x(t) \rangle$ of the TFIC with an initial domain-wall state within the same semi-classical approach.

When the field $h$ is tuned to its critical value $h_c=1$, the behaviour at the edge is described by Eq.~\eqref{eq:edge-crit} and, from the continuity equation, one eventually finds
\be
\begin{split}
&u(v\to \pm v_{max}^\mp)=u(\pm\infty) \\
&\pm \frac{\beta_r-\beta_l}{2 \pi} v_{max}^2 \left[ \mbox{arccos}\left(\frac{|v|}{v_{max}}\right) -\frac{|v|}{v_{max}}\sqrt{1-\frac{v^2}{v_{max}^2}}\right] + \\ 
&+ {\cal O}((v_{max}-|v|)^{5/2})
\end{split}
\label{arccos-crit}
\ee
instead of Eq.~\eqref{arcos}. 
Equation~\eqref{arccos-crit} is compared in Fig.~\ref{fig:edge2}(b) with the actual energy obtained from Eq.~\eqref{finalresultenergy}.  Their qualitative behaviour upon approaching the edge $v\simeq v_{max}$ is markedly different from the one  reported in panel (a) of the same figure for the non-critical chain.

\subsection{Correlation function and density of quasi-particles}
\label{subsec:corr-dens}

The procedure outlined above for predicting the evolution of the
energy current and density in the space-time scaling
limit can be extended to other relevant quantities. Here we focus on the
quasi-particle excitations and we study their dynamics along the chain by determining the corresponding spatial density and current. In particular, we consider the Fourier transform of the operator $\Psi_R(k)$, i.e.,
\be  
\gamma_x = \int_{-\pi}^{\pi} \frac{dk}{\sqrt{2\pi}}e^{ikx} \; \Psi_R(k).
\label{newfermion}
\ee
In terms of $\gamma_x$,  the total number operator $\hat{N}$ of quasi-particles excitations along the chain can be easily expressed as
\be
\hat{N} = \int_{-\pi}^{\pi} dk \; \Psi_R^\dagger(k) \Psi_R(k) =  \sum_{x=-\infty}^{+\infty} \gamma_x^\dagger \gamma_x ,
\label{ftrasformdensity}
\ee
which is conserved by the post-quench dynamics dictated by Eq.~\eqref{postquenchhamiltonian} since the 
quasi-particles $\gamma_x$ propagate freely along the chain, without experiencing scattering. 
According to Eq.~\eqref{ftrasformdensity},  $\gamma_x^\dagger \gamma_x$ can be interpreted as the density of quasi-particles on the lattice. In order to calculate the evolution of its expectation value, we first consider 
the two-point, equal-time correlation function $\langle \gamma_x^\dagger (t) \gamma_y (t) \rangle = \mbox{Tr}[\rho_0 \gamma_x^\dagger (t) \gamma_y (t) ]$ of the fermionic operators $\gamma_x$ and $\gamma_y$ at two distinct position $x$ and $y$ along the chain.   
The calculation of this quantity proceeds exactly as described above for the energy current and density in the space-time scaling limit and, for brevity, we do not reproduce it here but we report only the final expression:
\be
\begin{split}
\langle \gamma_x^\dagger (t) \gamma_y (t) \rangle = \int_{-\pi}^{\pi} \frac{dk}{2 \pi} e^{ik(x-y)} \Theta(v_g(k)t -\frac{x+y}{2}) f_{\beta_l}(k)&\\
                                  +\int_{-\pi}^{\pi} \frac{dk}{2 \pi}e^{ik(x-y)} \Theta(\frac{x+y}{2} -v_g(k)t)f_{\beta_r}(k).&
\end{split}
\label{correlations}
\ee
The density $n(x,t) = \langle \gamma_x^\dagger (t) \gamma_x (t) \rangle$ of these quasi-particle excitations  can be readily obtained by setting $y=x$ in the previous expression: 
\be
\begin{split}
n(x,t) = \int_{-\pi}^{\pi} \frac{dk}{2 \pi}& \left[ \Theta (v_g(k)t -x) f_{\beta_l}(k) \right.  \\ 
     &  \left. + \Theta(x -v_g(k)t)f_{\beta_r}(k) \right],
\label{finalparticle}
\end{split}
\ee
which, in the stationary limit $t\to\infty$, agrees with Eq.~(5) of Ref.~\cite{1367-2630-16-3-033028} for the mode occupation numbers in the NESS of the TFIC generated as discussed here.
Equation~\eqref{correlations} for $t\to\infty$, instead, has a simpler structure compared to the analogous expression which was derived in  Ref.~\cite{1367-2630-16-3-033028} (see Eq.~(7) therein) for the two-point correlation function.
This is due to the fact that Eqs.~\eqref{correlations} and \eqref{finalparticle} refer to the fermionic operator $\gamma_x$ in Eq.~\eqref{newfermion}, while Ref.~\cite{1367-2630-16-3-033028} provides the corresponding expression for the correlation function of the operators  $c_{x,y}(t)$ introduced with the Jordan-Wigner transformation in Eq.~\eqref{JordanWigner}, which the $\gamma_x$'s are linearly related to.         
Note that also the expression in Eq.~\eqref{finalparticle} for $n(x,t)$ can be given a semi-classical interpretation, analogous to the one explained in Fig.~\ref{fig:lightcone}. 
In fact, at $t=0$ one expects the fermionic excitations with momentum distribution $f_{\beta_r} (k)$ to be uniformly distributed in space on the right part of the chain, i.e., at $x>0$, while those with distribution $f_{\beta_l}(k)$ to be on the left one at $x<0$. (This initial condition is correctly reproduced by Eq.~\eqref{finalparticle} upon setting $t=0$.)
For $t>0$, the particles with a certain $k$ propagate ballistically and independently with their characteristic velocity $v_g(k)$ and therefore the ensemble of particles with initial statistics $f_{\beta_l}(k)$ reach all the points with $x<v_g(k) t$, while those with statistics  $f_{\beta_r}(k)$, all the points with $x> v_g(k) t$. Translated into equations, this picture yields directly Eq.~\eqref{finalparticle}.

Proceeding as done above for the energy density, a particle current $J_N (x,t)$ can be associated with $n(x,t)$ on the basis of a continuity equation in which the transported quantity is now the number of quasi-particles. Taking into account the  boundary condition $J_N (\pm \infty,t) = 0$, one eventually finds
\begin{equation}
J_N(x,t) = -\int_{-\infty}^{x} dx'  \; \frac{\partial n(x',t)}{\partial t}
\end{equation}
and, by using Eq.~\eqref{finalparticle}, 
\begin{equation}
J_N(x,t) = \int_{-\pi}^{\pi} \frac{dk}{2 \pi} v_g(k)\left[f_{\beta_l}(k)-f_{\beta_r}(k)\right]\Theta(v_g(k)t-x). 
\label{finalJn}
\end{equation}
This expression has exactly the same interpretation as Eq.~\eqref{finalresult} and, after having restricted the integration domain to $0\le k \le \pi$ in both expressions, it can be shown to coincide with the latter 
[see Eq.~\eqref{finalresultapp-2}] upon replacing $\varepsilon(k)$ (i.e., the energy) with 1, as appropriate for a quasi-particle density.
The plots of the scaling functions $J_N(x,t) \equiv J_N(v=x/t)$ and $n(x,t) \equiv n(v=x/t)$ in Eq.~(\ref{finalparticle}) display the same qualitative features as the energy current and energy density in Fig.~\ref{fig:scaling}, respectively, with a marked propagating front which moves ballistically. In fact, by 
looking specifically at Fig.~\ref{fig:scaling}, one realizes that as time elapses, the gradient of the particle number $n(x,t)$ around the junction point $x=0$ decreases and vanishes asymptotically as $t \rightarrow +\infty$. Since the particle current $J_N$ keeps anyhow a non-zero value, it cannot be proportional to the gradient of $n(x,t)$, as required by diffusion, but to $n(x,t)$, as expected by ballistic transport. Accordingly, every possible diffusive component of the current $J_N$ gets suppressed, making the steady-state transport of fermions purely ballistic. 

\begin{figure}
\begin{center}
\includegraphics[width=0.8\columnwidth]{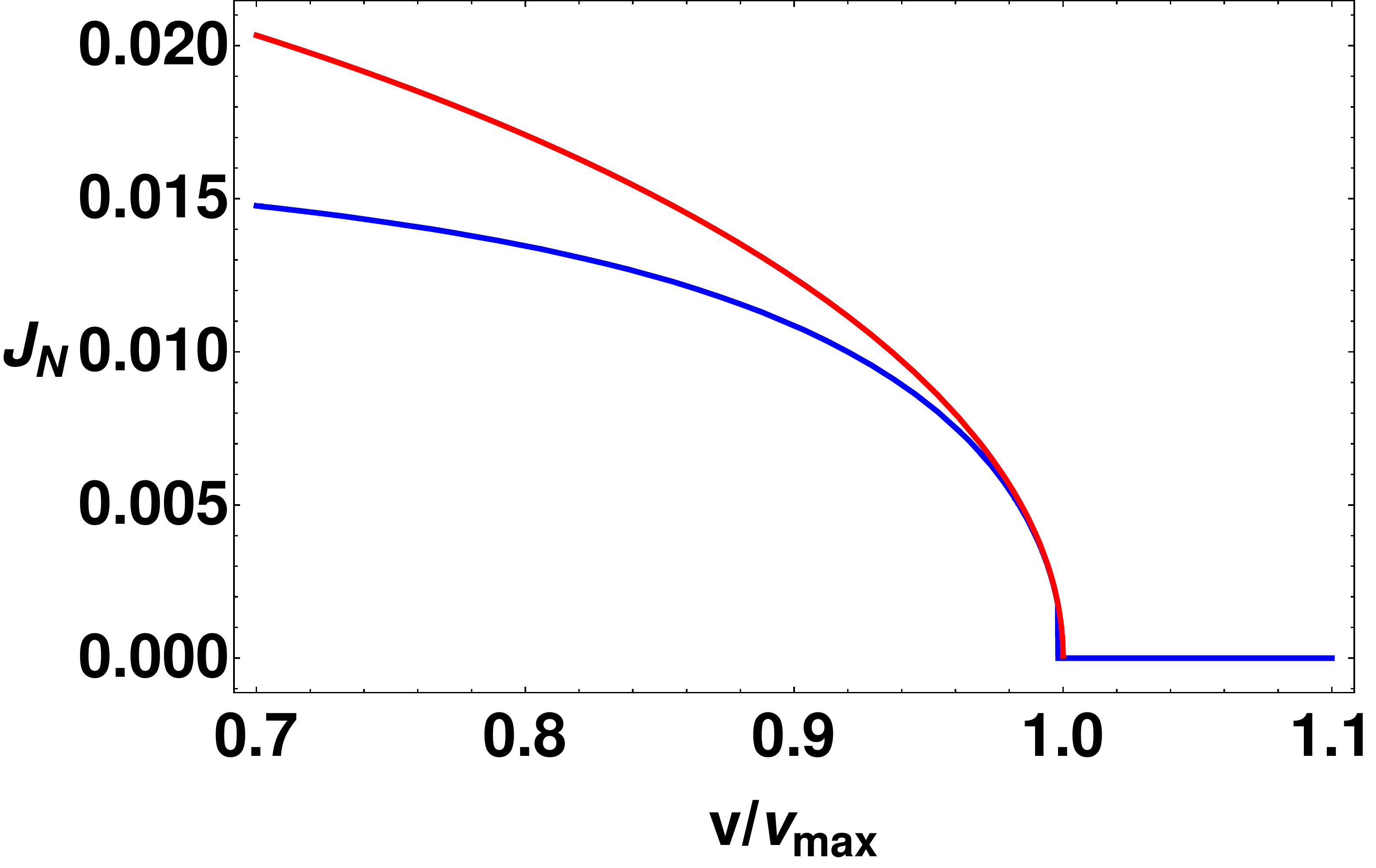}\\[-1mm]	
(a)\\[2mm]
\includegraphics[width=0.8\columnwidth]{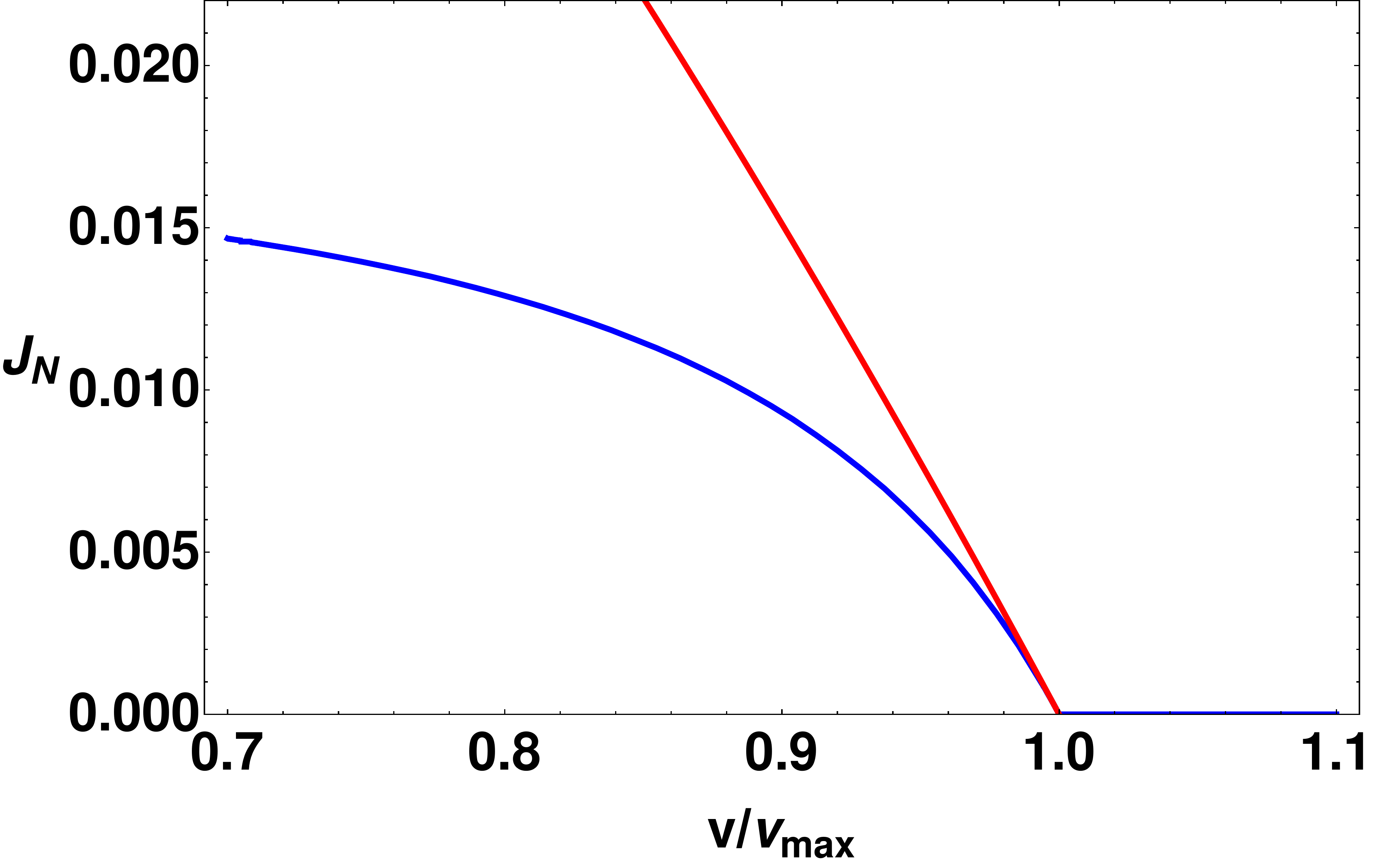}\\[-1mm]
(b)
\end{center}	
\caption{Fermion current $J_N(v)$ as a function of the scaling variable $v/v_{max}$ for a TFIC with (a) $h=1.2$ and (b) $h=h_c=1$, prepared in an initial state with $\beta_l=2$ and $\beta_r=3$ ($J=1$). The lower (blue) curves correspond to the analytic expression discussed in the main text while the upper (red) curves to its approximations close to the edge $v \rightarrow v_{max}^- = Jh$.}
\label{fig:edge-JN}
\end{figure}

\begin{figure}
\begin{center}
\includegraphics[width=0.8\columnwidth]{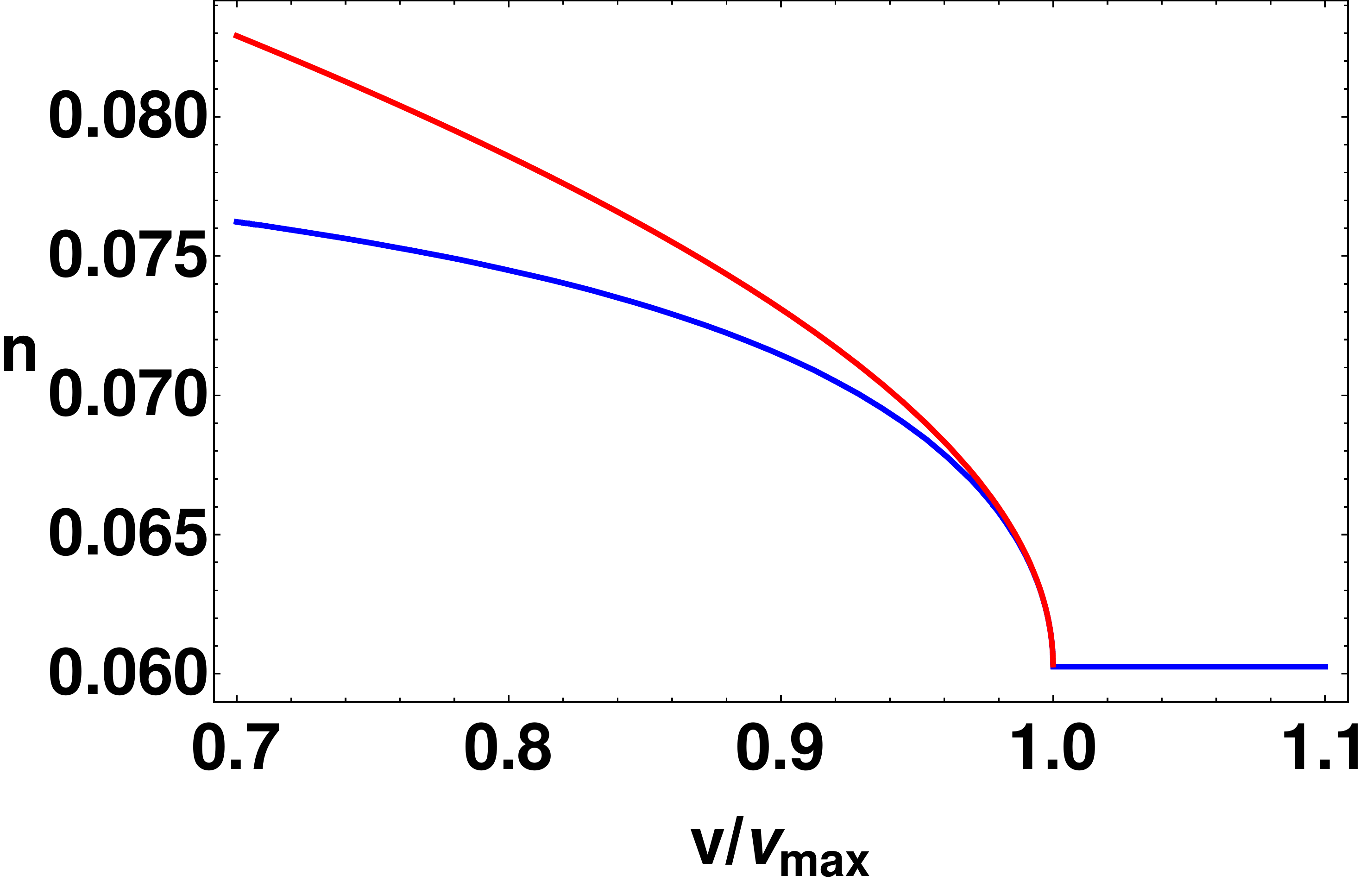}\\[-1mm]	
(a)\\[2mm]
\includegraphics[width=0.8\columnwidth]{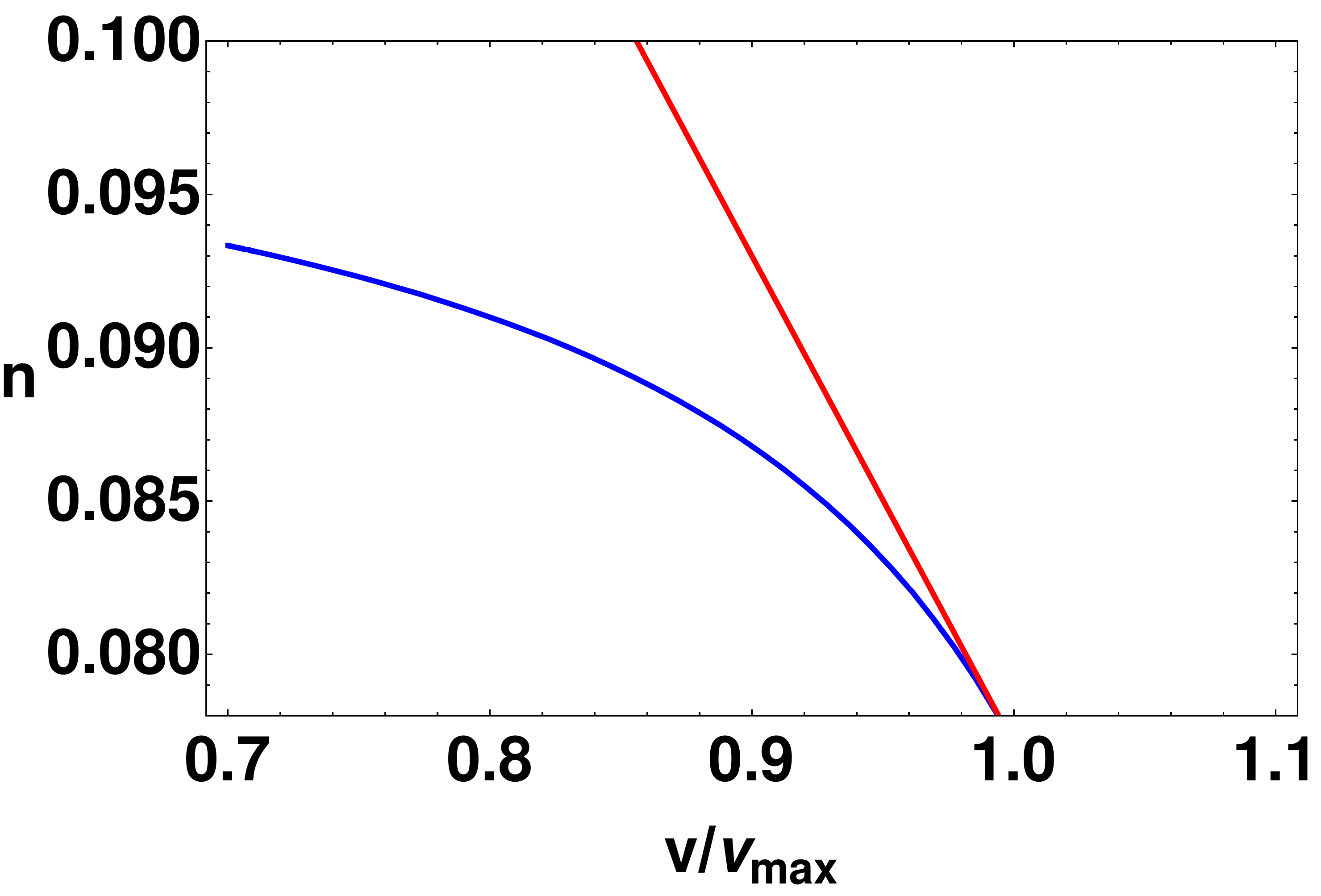}\\
(b)
\end{center}	
\caption{Fermion concentration $n(v)$ as a function of the scaling variable $v/v_{max}$ for the same TFIC as in Fig.~\ref{fig:edge-JN}. The lower (blue) curves correspond to the analytic expression discussed in the main text for (a) a non-critical and (b) the critical value of the transverse field $h$, while the upper (red) curves to the corresponding  approximations close to the edge $v \rightarrow v_{max}^- = Jh$.}
\label{fig:edge-N}
\end{figure}

As argued above and discussed in more detail in Appendix~\ref{appendix2}, 
$J_E(v)$ and $J_N(v)$ have analogous forms and actually they can be obtained from the function $J_a(v)$ introduced in Eq.~\eqref{eqapp:Ja} by setting $a=1$ and $0$, respectively. Accordingly, $J_N(v)$ has the same expression as Eqs.~\eqref{rl} and \eqref{integration} with  ${\cal J}_1$ replaced by ${\cal J}_0$ 
\be
\begin{split}
{\cal J}_0(\beta,v)= &\frac{1}{2 \pi \beta}\left\{ G_0(\beta [\varepsilon_>(v)-\varepsilon_<(v)])+\right. \\
                    & \qquad \left. -G_0(\beta [\varepsilon_>(v)+\varepsilon_<(v)])\right\},
\label{eq:integration2}
\end{split}
\ee
and $G_1$ by
\be
G_0(x) = \ln\left( 1+e^{-x}\right).
\label{eq:G2}
\ee
In particular, also $J_N(v)$ shows a non-analytic behaviour as $v$ approaches  $\pm v_{max}$, which has the same 
form as Eq.~\eqref{semicircular}, with $C_1$ replaced by the constant $C_0$ reported in Eq.~\eqref{eq:genexprC}. Moreover, similarly to $J_E(v)$, the qualitative features of the approach of $J_N$ to the edge do change if the transverse field $h$ is tuned to its critical value $h_c=1$, with
\be
\begin{split}
J_N(v) =  & \frac{(\beta_r-\beta_l)}{4 \pi}(v_{max}^2-v^2) 
+ {\cal O}((v_{max}-|v|)^{2}).
\label{eq:edgecritical2}
\end{split}
\ee
In Fig.~\ref{fig:edge-JN} we compare the actual approach of $J_N(v)$ to the edge $v_{max}$ with the approximation provided by the semicircular law for the non-critical case in panel (a) and by Eq.~\eqref{eq:edgecritical2} for the critical case in panel (b).
From the continuity equation relating the particle current $J_N(v)$ with the fermionic density $n(v)$, exploiting the edge behaviour of Eq.~\eqref{semicircular} with $C_1$ replaced by $C_0$, one can determine the behaviour of $n(v)$ upon approaching the edge, based on the previous results. For $h \neq 1$, it turns out to be
\be
\begin{split}
n(v\to \pm v_{max}^\mp)=n(\pm\infty)\pm C_0 & \; \mbox{arccos}\left(\frac{|v|}{v_{max}}\right) \\ + &{\cal O}((v_{max}-|v|)) ,
\label{eq:nedge}
\end{split}
\ee
where $n(\pm\infty)$ has, for the fermionic density, the same meaning as $u(\pm\infty)$ in Eq.~\eqref{arcos} for the energy density. When $h$ is tuned to its critical value $h_c=1$, the previous equation must be corrected taking into account Eq.~\eqref{eq:edgecritical2}: 
\be
\begin{split}
n(v\to \pm v_{max}^\mp)=n(\pm\infty) \mp &\frac{\beta_r - \beta_l}{2 \pi} (|v|-v_{max})\\ 
&+{\cal O}((v_{max}-|v|)^{2}).
\label{eq:nedge_critical}
\end{split}
\ee
In Fig.~\ref{fig:edge-N} the behavior of $n(v)$ as $v$ approaches the edge $v_{max}$ is compared with that of the expansion provided by Eq.~\eqref{eq:nedge} for the non-critical case in panel (a), and by Eq.~\eqref{eq:nedge_critical}
for the critical one in panel (b).
%
%
%
%
%
%

Analogously to the case of the energy current discussed above, the value $J_N^{NESS}$ of the particle current $J_N$ in the steady state can be exactly calculated by setting $v=0$ in the expression obtained by modifying Eqs.~\eqref{rl} and \eqref{integration} as mentioned above, with $G_1 \mapsto G_0$, see Eq.~\eqref{eq:G2}; this result into Eq.~\eqref{eq:JENESS-here} with $G_1 \mapsto G_0$, i.e.,
\be
J_N^{NESS} (\beta_l,\beta_r)  = {\cal J}_0(\beta_l,0) - {\cal J}_0(\beta_r,0),
\label{eq:JNNESS-here}
\ee
where
\be
{\cal J}_0(\beta,0) =  \frac{1}{2\pi \beta} \ln \left( \frac{1+e^{-\beta J|h-1|}}{1+e^{-\beta J(h+1)}}\right);
\ee
this expression, as well as the one of the energy current in Eq.~\eqref{eq:JENESS-here}, takes the general form of Eq.~\eqref{vitiresult}, which often appears in studies of transport in non-equilibrium quantum stationary states, see, e.g., Refs.~\cite{collura2014quantum,collura2014non,de2013nonequilibrium}.
In the case of a chain with critical transverse field $h=h_c=1$,  
in the limit $\beta_{r,l}J \gg 1$, one readily finds 
\begin{equation}
J_N^{NESS} = \frac{\mbox{ln}2}{2 \pi} \; (\beta_l^{-1} - \beta_r^{-1}),
\end{equation}
which coincides with the result obtained in Refs.~\cite{collura2014quantum,collura2014non} for a local quench of non-interacting Fermi gases in one spatial dimension.

\section{Correlations and energy current near the edge} 
\label{thirdsection}

In this section we investigate in more detail the dynamics of the propagating front near the edge, i.e., for  $x \simeq \pm v_{max} t$: in fact, experience with analogous cases \cite{eisler2013full,hunyadi2004dynamic} suggests that both the correlations and the energy current acquire non-trivial corrections within a distance $\Delta x \propto t^{1/3}$ from the edge and we shall see below how they emerge in the present setting. As the relative width $\Delta x/x \sim t^{-2/3}$ of the spatial region interested by these corrections vanishes in the space-time scaling limit, these features are not captured by the previous analysis and therefore they require a separate treatment.

\subsection{Correlation functions}

\begin{figure}
	\centering
	\includegraphics[width=1\columnwidth]{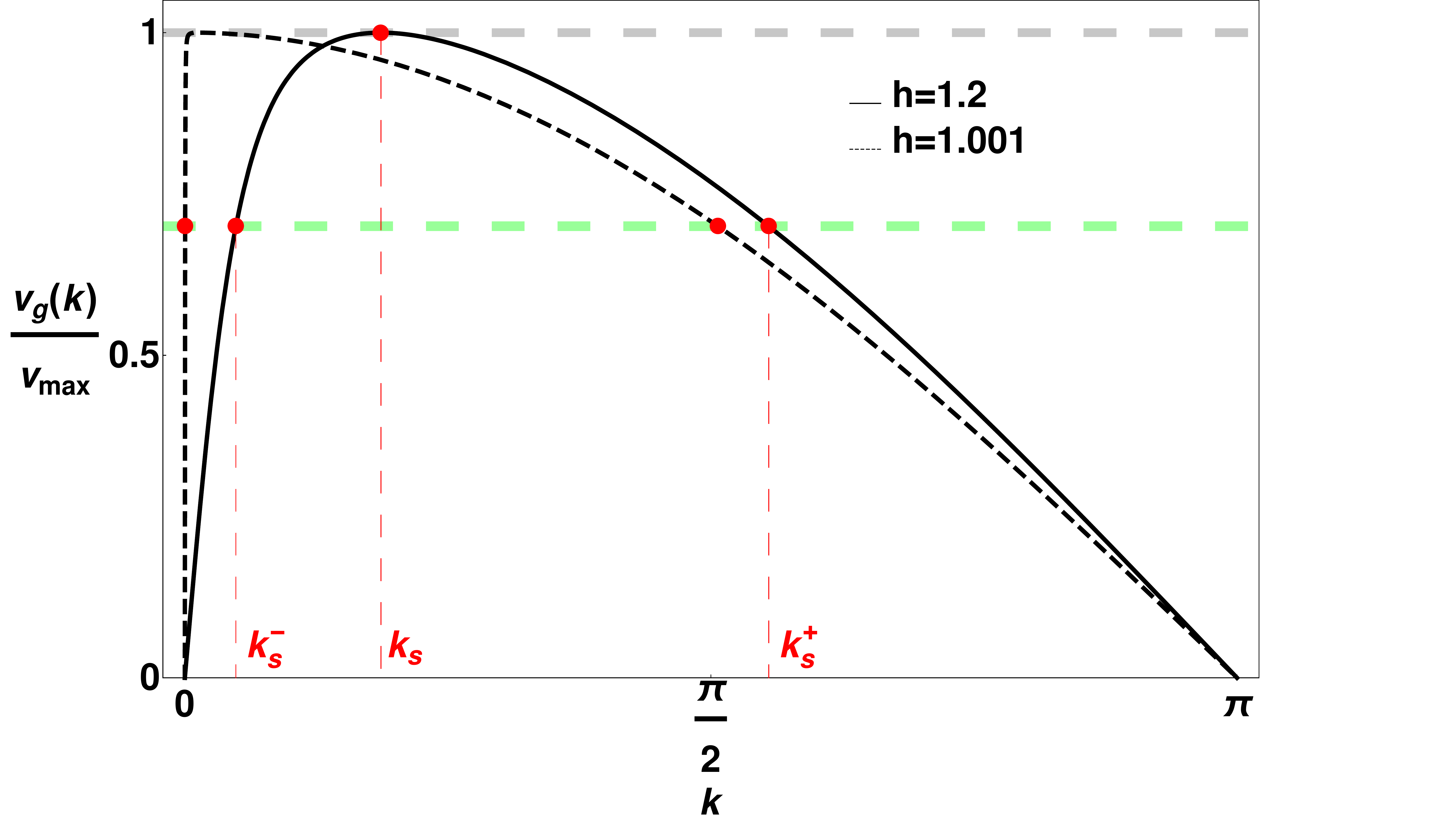}		
	\caption{Graphical representation of the solution of Eq.~\eqref{stationaryphase}. The dashed and solid curves represent the group velocity $v_g(k)$ as a function of $k$ for $h=1.001 \simeq h_c$ and $h=1.2$, 
respectively, with $J=1$. The lowest (green) horizontal dashed line indicates a certain assigned value of $|v=x/t|$ and, correspondingly, the values $k_s^\pm(v)$ of $k$ at which it crosses the previous curves are the solution of the equation. As $|v|$ approaches $v_{max}$, indicated by the upper dashed horizontal line, the two solutions $k_s^+$ and $k_s^-$ merge into a unique value $k_s$.} 
	\label{fig:timedensity}
\end{figure}

\begin{figure}
	\centering
	\includegraphics[width=0.8\columnwidth]{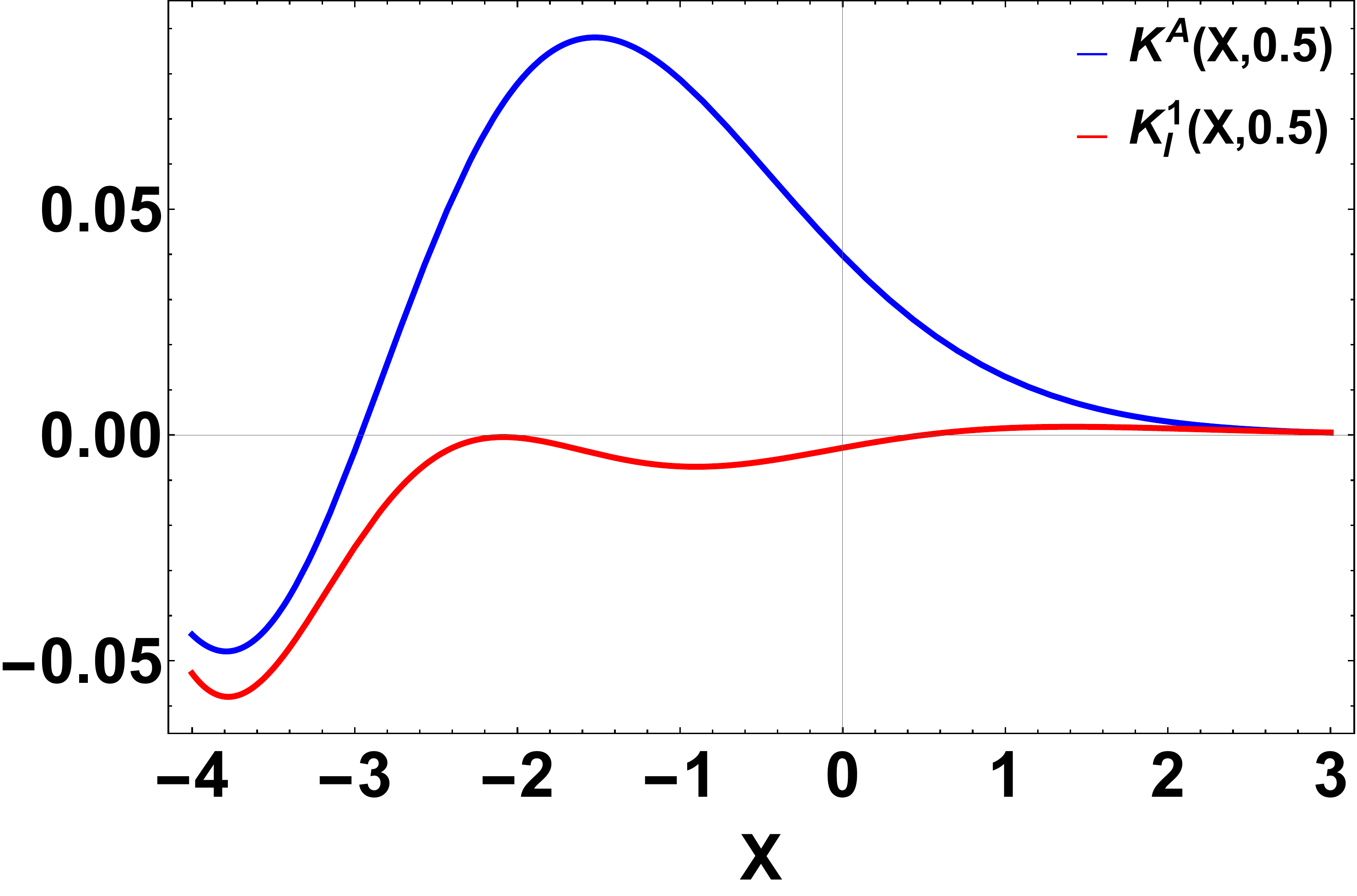}\\
	\includegraphics[width=0.80\columnwidth]{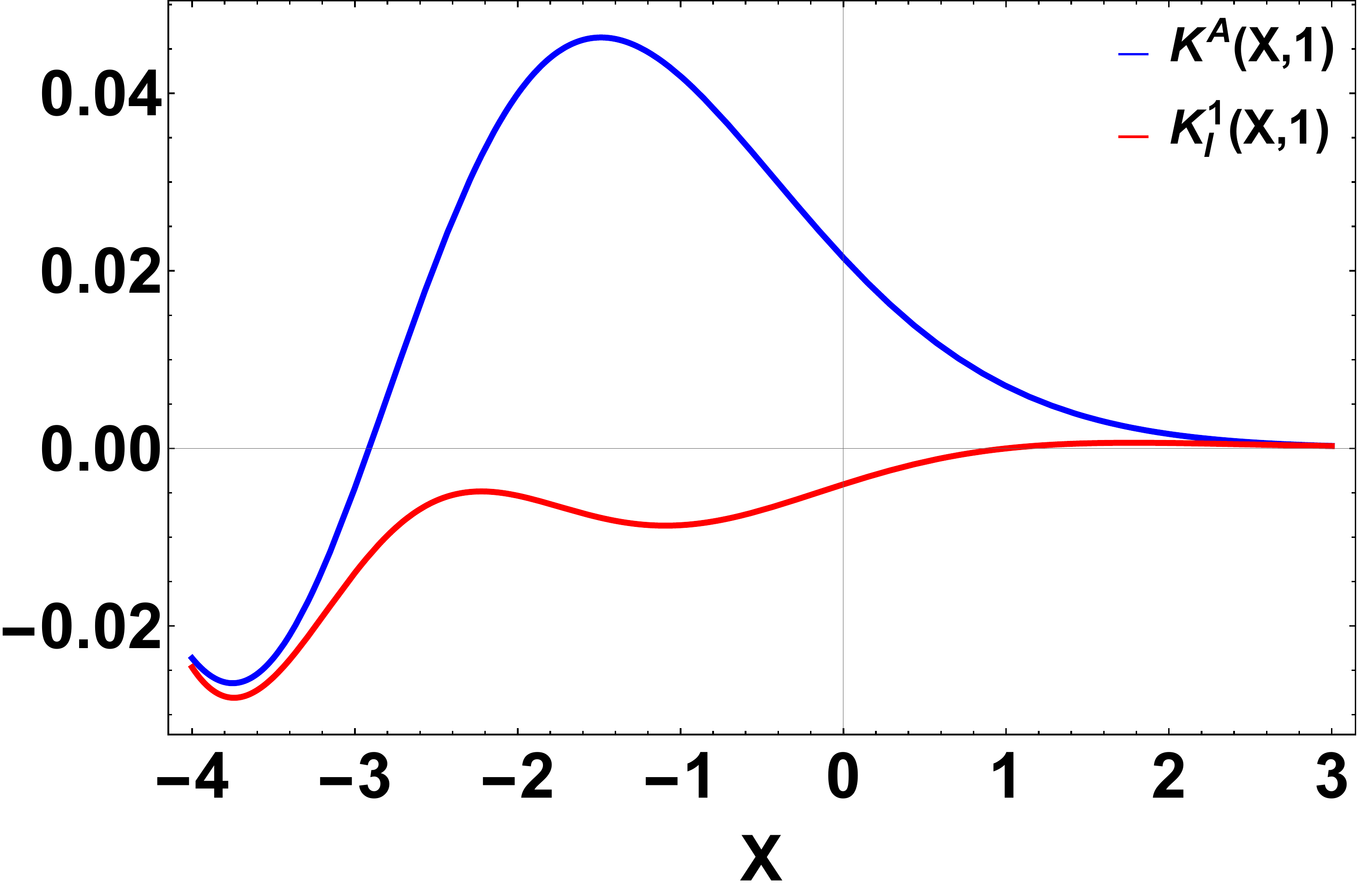}\\	
	\hspace{-1.5mm}\includegraphics[width=0.825\columnwidth]{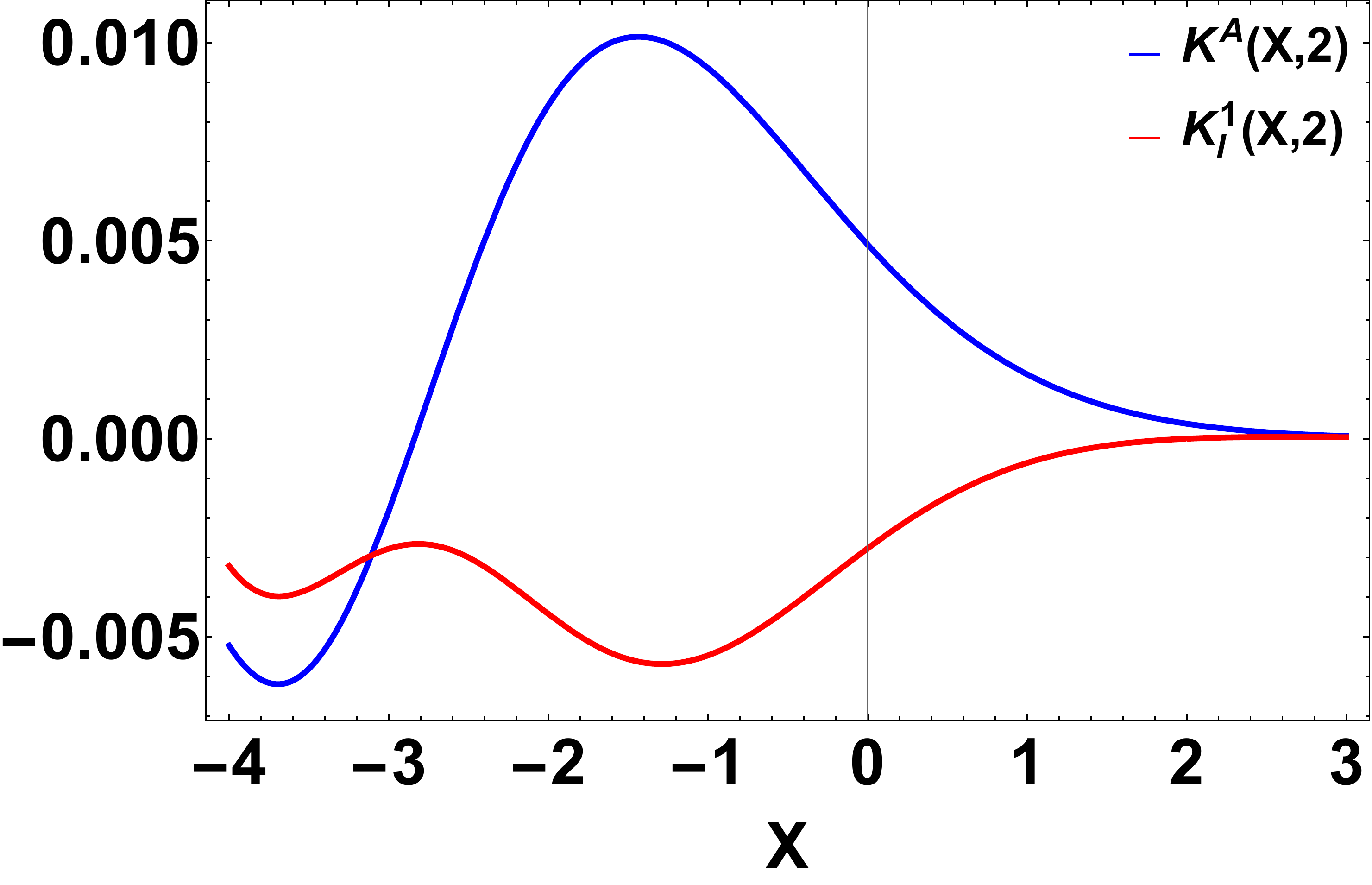}
	\caption{%
	Airy Kernel $K^{A}(X,Y)$ (blue) and first-order correction $(2/v_{max}t)^{1/3}K^1_{\beta_l}(X,Y)$ (red) as functions of $X$ for $Y = 0.5$, 1, and 2, from top to bottom. Here we assume $h=1.2$, $J=1$, and $t=100$.}	
	\label{fig:corrections}
\end{figure}

Consider the two-point correlation function
\be
\begin{split}
\langle \gamma_x^\dagger (t) \gamma_y (t) \rangle &= \int_{-\pi}^{\pi} \int_{-\pi}^{\pi} \frac{dk dk'}{2 \pi} e^{i[\phi_{x,t}(k)-\phi_{y,t}(k')]} I(k,k')  \\
 &= \langle \gamma_x^\dagger (t) \gamma_y (t) \rangle_{l} + \langle \gamma_x^\dagger (t) \gamma_y (t) \rangle_{r},  
 \end{split} 
\label{starting3}
\ee
where $\phi_{x,t}(k) \equiv \varepsilon(k)t-kx$ and $I(k,k')$ is defined in Eq.~(\ref{integralapp}). Note that the r.h.s.~of this equation naturally decomposes as the linear superposition of two distinct contributions $\langle \gamma_x^\dagger (t) \gamma_y (t) \rangle_{l,r}$  corresponding to the effect of considering separately one of the two half chains initially populated according to the corresponding thermal distribution, while the other unoccupied (i.e., with $f_{\beta_r,\beta_l} = 0$).
As discussed after Eq.~\eqref{operator2} and in Appendix \ref{appendix2}, the space-time scaling limit of expressions such as Eq.~\eqref{starting3} is conveniently studied after the change of variable $Q = k-k'$ and $K = (k+k')/2$ and after expanding around $Q=0$ up to first order in $Q$: this renders Eq.~\eqref{correlations} discussed in the previous section. 
Here, instead, we are interested in the behaviour of this quantity near the edge of the propagating front, corresponding to having $|x|, |y| \simeq v_{max} t$: in this case, higher-order corrections in the expansion of the phases $\phi_{x,t}(k)$ and $\phi_{y,t}(k')$ around the respective stationary points become important and therefore they have to be accounted for. 
Namely, as $v \equiv x/t$ approaches $\pm v_{max}$, the two solutions $k_s^+(v)$ and $k_s^-(v)$ of the stationary phase equation (see Appendix~\ref{appendix2} for details)
\be
v_g(k_s^\pm(v)) = |v|
\label{stationaryphase}
\ee
merge into a unique stationary point $k_s = k_s^\pm(v_{max})$  obtained by taking $v=v_{max}$ into Eq.~\eqref{criticalks}, at which the second derivative of the phases $\phi_{x,t}(k)$ and $\phi_{y,t}(k')$ vanishes, as shown in Fig.~\ref{fig:timedensity}. 
Accordingly, one expects non-trivial corrections due to higher-order terms in the expansion of $\phi_{x,t}(k)$ and $\phi_{y,t}(k')$ around the stationary point $k_s$. In particular,  the third-order correction is expected to provide the dominant contribution and, accordingly, the phase is approximated as
\be
\label{phaseexpansion}
\begin{split}
&\phi_{x,t}(k) =  \varepsilon(k_s)t -k_s x + (k-k_s)(v_{max}t-x) \\ 
               & \qquad\qquad -\frac{v_{max}}{3!}(k-k_s)^3 t + {\cal O}((k-k_s)^4),
\end{split}
\ee
where we used the fact that $\varepsilon'''(k_s) = -v_{max}$ and $\varepsilon''(k_s)=0$, see Eq.~\eqref{eq:app-exp-phi}. 
In order to evaluate the integral in Eq.~\eqref{starting3} within this approximation, it is convenient to introduce the variables $K= (v_{max}t/ 2)^{1/3}(k-k_s)$ and $Q= (v_{max}t/2)^{1/3}(k'-k_s)$ instead of $k$ and $k'$: by expanding the Fermi-Dirac distributions $f_{\beta_{l,r}}$ which, via $I(k,k')$ [see Eq.~\eqref{integralapp}], appears on the r.h.s.~of Eq.~\eqref{starting3}, one obtains an expansion of the "left" contribution $\langle \gamma_x^\dagger (t) \gamma_y (t) \rangle_l$  to the correlation function in Eq.~\eqref{starting3} (an analogous result holds for the "right" contribution $\langle \gamma_x^\dagger (t) \gamma_y (t) \rangle_r$). Up to order $t^{-2/3}$, one finds
\be
\begin{split}
& \frac{{\langle \gamma_x^\dagger (t) \gamma_y (t) \rangle}_l}{e^{-ik_s(x-y)}f_{\beta_l}(k_s)} =\left(\frac{2}{v_{max}t}\right)^{1/3} K^{A}(\mbox{X},\mbox{Y}) \\ 
             & \qquad\qquad+ \left(\frac{2}{v_{max}t}\right)^{2/3} K^{1}_{\beta_l}(\mbox{X},\mbox{Y}) + {\cal O}\left(\frac{1}{v_{max}t}\right) 
\label{airyexpansion}
\end{split} 
\ee
where the relevant scaling variables are 
\be
X \equiv \frac{x-v_{max}t}{(v_{max}t/2)^{1/3}}, \quad \mbox{and} \quad
Y \equiv  \frac{y-v_{max}t}{(v_{max}t/2)^{1/3}}.
\label{scalingvariables} 
\ee
The analogous expression for $\langle \gamma^\dagger_x(t)\gamma_y(t)\rangle_r$ is obtained from the previous one upon replacing $\beta_l$ with $\beta_r$. 
The leading contribution on the r.h.s.~of Eq.~\eqref{airyexpansion} is the so-called Airy kernel, which is given by
\be
K^{A}(X,Y) = \frac{\mbox{Ai}(X)\mbox{Ai'}(Y)-\mbox{Ai'}(X)\mbox{Ai}(Y)}{X-Y}, 
\label{AiryKernel}
\ee
in terms of the Airy function $\mbox{Ai}(X)$ (see Eq.~\eqref{eq:def-Airy} for its integral representation); as Eq.~\eqref{airyexpansion}  depends on $x$, $y$, and $t$ via the scaling variables in Eq.~\eqref{scalingvariables}, it expresses the scaling behaviour of the correlation function, which emerges within a spatial region of thickness $\propto t^{1/3}$ from the location $\propto t$ of the edge and therefore it cannot be captured by the semi-classical, hydrodynamic, approach discussed in the previous Section, as anticipated above. 
In particular the Airy kernel results from the fact that the two solutions of Eq.~\eqref{stationaryphase} coincide 
near the edge of the light-cone and, in fact, this kernel emerges rather generically in the literature concerning
free spinless fermionic chains \cite{allegra2016inhomogeneous,eisler2013full}, where it has been reported for the case of an initial state consisting of a fully occupied half chain and   for a more general initial factorized Fermi 
sea state \cite{viti2015inhomogeneous}. 
Equation~\eqref{airyexpansion} shows that the leading effect of an initial state with two different (finite) temperatures 
$\beta_l^{-1}$ and $\beta_r^{-1}$ is the presence of the corresponding distributions $f_{\beta_l}(k_s)$ [and $f_{\beta_r}(k_s)$ in the analogous expression for $\langle \gamma^\dagger_x(t)\gamma_y(t)\rangle_r$ which we do not report here]
as a multiplicative factor of the Airy scaling function which emerges also for the different initial conditions mentioned above, corresponding to $f_{\beta_l}(k) = 1$ and $f_{\beta_r}(k) = 0$ for the domain wall initial state, and to $f_{\beta}(k) = \Theta(k_F - k)$ for the Fermi sea state where $k_F$ denotes the Fermi momentum.
In passing we mention that the Airy kernel and a generalization of it emerge at the spatial edge of a system at zero and finite temperature, respectively, also in the case of a one-dimensional gas of free fermions confined by an harmonic potential \cite{eisler2013universality,dean2015finite,dean2016non}. However, in this case, the edge does not expand in time but is rather fixed by the presence of the harmonic potential which makes the fermion density vanish beyond a certain distance from the center of the trap. Close to that edge, 
the correlation function is expressed as a determinantal process whose kernel can be interpreted as an extension of the Airy one. 
In the present case of  two Ising chains with an initial thermal distribution but with two different temperatures $\beta_l^{-1}$ and $\beta_r^{-1}$, the leading-order behaviour at the edge is modified in a somehow expected way, i.e., it has the same form as in the aforementioned cases with domain wall and factorized Fermi sea  
except for the multiplicative factor determined by the corresponding thermal distribution $f_{\beta_{r,l}}(k_s)$ evaluated at the saddle point $k_s$.
In order to highlight the effects of the first non-trivial contributions due to the finite initial temperatures, we report in Eq.~(\ref{airyexpansion}) also the rescaled first-order correction $(2/v_{max}t)^{1/3} K_{\beta_l}^1$ [see its explicit expression in Eq.~(\ref{kernelresult})], which is compared with the leading order $K^A$ in Fig.~\ref{fig:corrections}.
One can notice that, close to $X=0$, i.e., for $|x| \simeq v_{max}t$, the first-order correction $K_{\beta_l}^1$ turns out to be  one order of magnitude smaller than the Airy Kernel $K^A$, for both $Y=0.5$ and $1$. For $Y=2$ and $t=100$, as in Fig.~\ref{fig:corrections}, instead, the two contributions become comparable; as a matter of fact as $Y$ increases one gets progressively away from the edge region for the $Y$ variable and thus correlations are expected to be captured in a less precise way by the expansion in Eq.~(\ref{airyexpansion}). 

\subsection{Energy current}

The procedure described above for studying the correlation functions close to the edge, can be also applied to the energy current. 
In this case, at the leading order for $x \simeq v_{max}t$ (see Appendix \ref{appendix3} for details, with an analogous expression holding for $x \simeq -v_{max}t$ due to the symmetry $J_E(x,t) = J_E(-x,t)$), one finds
\begin{equation}
J_E(X,t) = 
\varepsilon(k_s) v_{max}\left[ n_{l}(X,t)-n_{r}(X,t)\right],
\label{eisler}
\end{equation}
where the scaling variable $X$ is given in Eq.~\eqref{scalingvariables}, 
\be
\begin{split}
n_{l,r} (X,t) &= \langle \gamma_x^\dagger (t) \gamma_x (t) \rangle_{l,r} \\
&= (v_{max}t/2)^{-1/3} f_{\beta_{l,r}}(k_s)K^{A}(X,X),
\end{split}
\label{eq:nalpha}
\ee
(see Eq.~\eqref{starting3})
and the kernel 
\be
K^A(X,X) = \left[\mbox{Ai}'(X)\right]^2 -X\left[\mbox{Ai}(X)\right]^2
\label{eq:KAXX-def}
\ee 
is obtained as the limit $Y\to X$ of Eq.~\eqref{AiryKernel}.
As it is clear from the structure of Eq.~\eqref{eisler}, the current $J_E$ at the leading order can be given a simple interpretation as resulting from the superposition of the energy current $\varepsilon(k_s) v_{max} n_{l}(X,t)$ close to the edge, due to the fastest excitations (at temperature $\beta^{-1}_l$) propagating rightward and originally produced on the left part of the chain and the one with opposite sign $-\varepsilon(k_s) v_{max} n_{r}(X,t)$ due to those originating on the right part of the chain and propagating leftward. 
In particular, the semi-classical limit discussed in Sec.~\ref{sec:JEandu} ---
in which $v\equiv x/t \leq v_{max}$ is kept constant as $t\to\infty$ --- corresponds to having $X \propto (v-v_{max}) t^{2/3} \to -\infty$ here and, in fact, in this limit, Eq.~\eqref{eisler} renders the behavior of $J_E(v)$ close to the edge, reported in Eq.~\eqref{semicircular}. 
This can be easily seen by using Eq.~\eqref{C1constant} in Eqs.~\eqref{eisler} and \eqref{eq:nalpha} and by taking into account the asymptotic behaviour of the Airy Kernel $K^A(X,X) \rightarrow \sqrt{-X}/\pi$, which follows from Eq.~\eqref{Airy-inf}.

\begin{figure}
\centering
\includegraphics[width=0.82\columnwidth]{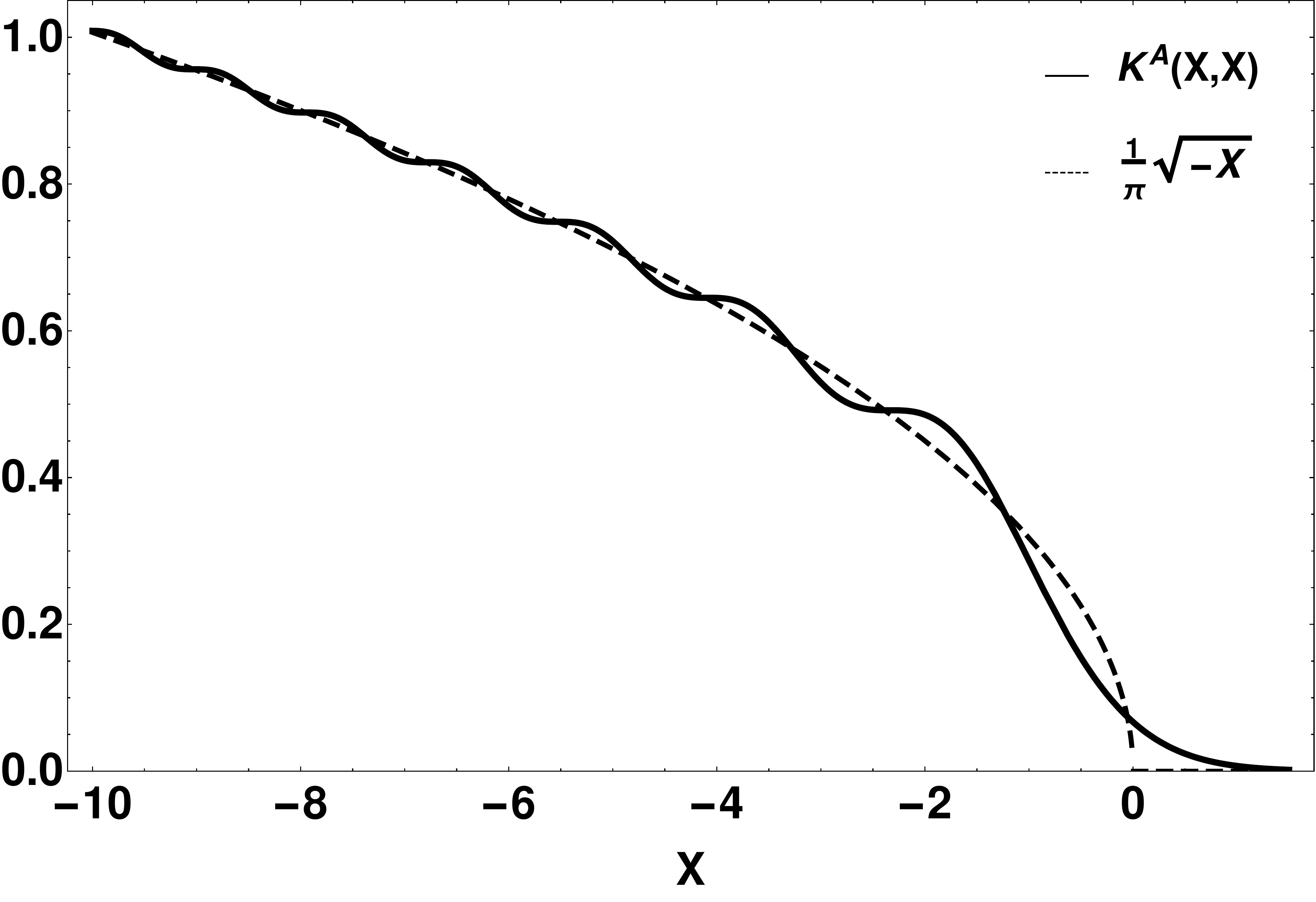}	
\caption{%
Dependence of the energy current $J_E(X,t)$ on the rescaled coordinate $X$ [see Eq.~\eqref{scalingvariables}] near the edge $X=0$, for $t=1$, $h=1.2$, $J=1$, $\beta_l=2$, and $\beta_r=3$. 
}
\label{fig:heatedge}
\end{figure}

\begin{figure}
	\centering
	\includegraphics[width=0.8\columnwidth]{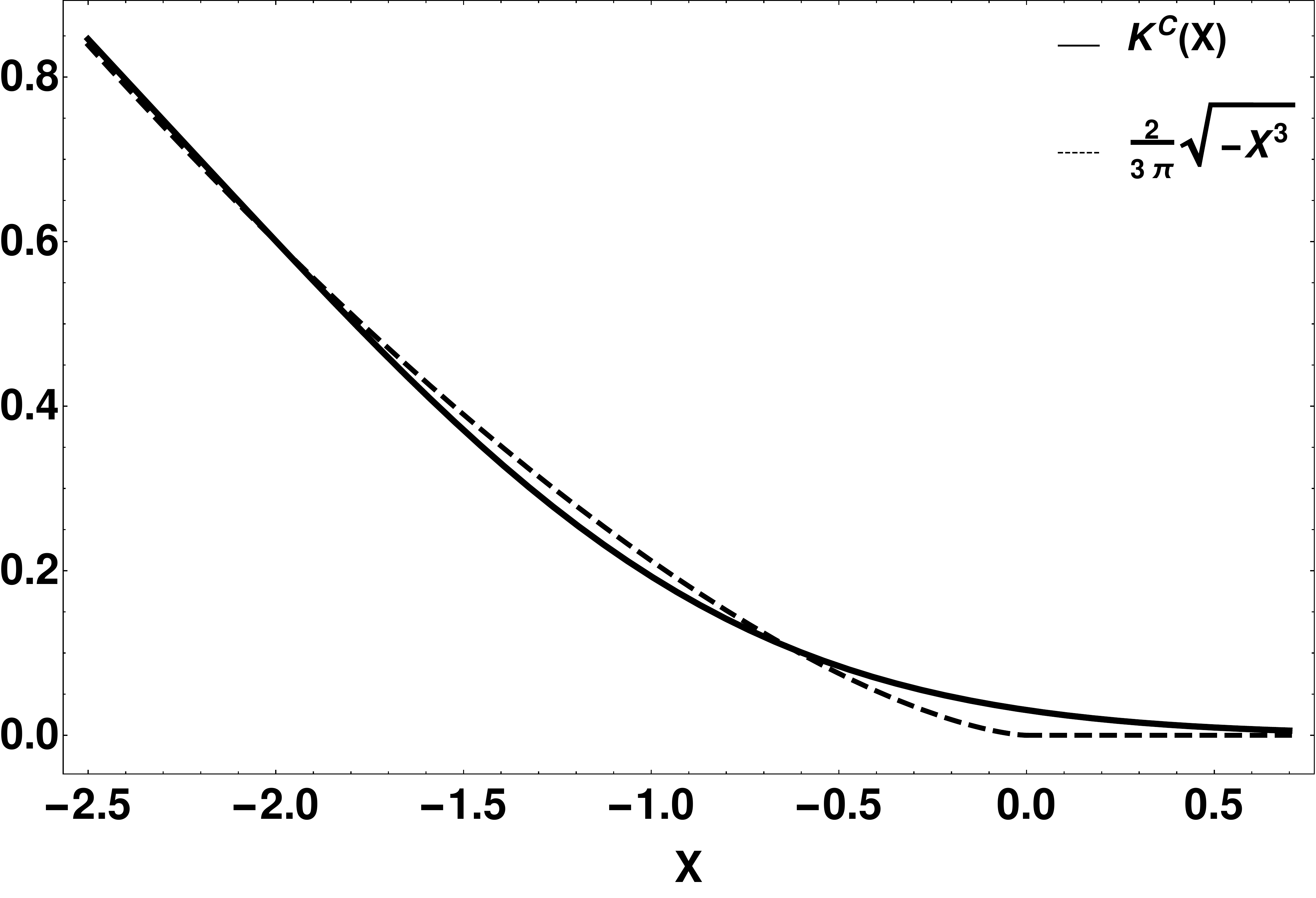}	
	\caption{%
Dependence of the energy current $J_E(X,t)$ on the rescaled coordinate $X$ [see Eq.~\eqref{eq:newX}] near the edge $X=0$, for a critical value $h=h_c=1$ of the transverse field. The remaining parameters are the same as those of Fig.~\ref{fig:heatedge}.
}
\label{fig:criticaledge}
\end{figure}

Figure \ref{fig:heatedge} presents a plot of the current $J_E(X,t) \times (v_{max}t/2)^{1/3}$ as a function of the scaling variable $X$, which is compared with Eq.~(\ref{eisler}), while the dashed line indicates the asymptotic behavior for $X\to-\infty$.
%
The solid line features the typical staircase behaviour caused by the cubic term in the expansion of the phase in Eq.~(\ref{phaseexpansion}) which is therefore not captured by the semi-classical limit discussed in Sec.~\ref{sec:JEandu}, which actually corresponds to the square-root envelope (dashed line) of the boundary scaling regime. These oscillations are similar to those obtained in Ref.~\cite{eisler2013full} for a free-fermionic chain starting from a domain-wall initial state, in which case the subsequent steps in the staircase have been explained on the basis of the correspondence existing between the counting statistics of free fermions at the edge of a propagating front and that one of the eigenvalues of a random matrix. 
As we noted in Sec.~\ref{sec:JEandu} (see, in particular, Fig.~\ref{fig:color}), the behavior of the current $J_E$ in the space-time scaling limit changes qualitatively when the transverse field $h$ takes its critical value $h_c=1$. Accordingly, one expects the edge behavior to be affected as well. In fact, it is straightforward to note that $k_s\to 0$ as $h \to 1$  (see Fig.~\ref{fig:timedensity} and Eq.~\eqref{criticalks}) and, correspondingly, $\varepsilon(k_s) \to 0$ in the same limit, which makes the expression for $J_E(X,t)$ in Eq.~(\ref{eisler}) vanish identically.
In this case, within the stationary-phase approximation adopted here, one has to keep terms up to the first non-vanishing order $\propto 1/t$ in the expansion in $k$ and $k'$ around $k_s$.  
Proceeding in this way (see Appendix \ref{appendix3} for details), one finds
\be
J_E(X,t) = \frac{1}{t}\frac{v_{max}^2(\beta_r - \beta_l)}{2} K^c(X)
\label{finalkernel}
\ee      
where $X$ in Eq.~\eqref{eisler} (see Eq.~\eqref{scalingvariables}) is replaced by
\be
X \equiv \frac{x - v_{max}t}{(v_{max} t/8)^{1/3}}
\label{eq:newX}
\ee
and 
\be
\begin{split}
K^c(X) = \frac{4}{3}\big\{X^2[\mbox{Ai}(X)]^2-\frac{1}{2}\mbox{Ai}(X)\mbox{Ai'}(X) &\\
          - X[\mbox{Ai'}(X)]^2\big\}. &
\label{kernelcritical}
\end{split}
\ee
By using the property of the Airy function 
\be
\mbox{Ai}''(X) = X \mbox{Ai}(X),
\label{eq:diff-eq-Airy}
\ee
one can easily show by differentiating the previous equation that $K^c(X)$ is related to $K^A(X,X)$ in Eq.~\eqref{eq:KAXX-def} by (see also Appendix~\ref{appendix3})
\be
K^c(X) = 2 \int_X^{+\infty} dY \, K^A(Y,Y).
\label{eq:rel-kernels}
\ee
As in the case $h\neq h_c$ discussed above, Eq.~\eqref{finalkernel} renders Eq.~\eqref{eq:edge-crit} after taking into account the asymptotic behaviour $K^c(X\to -\infty)\simeq 2\sqrt{-X^3}/(3\pi)$, which can be again obtained from Eq.~\eqref{Airy-inf}. 
Figure~\ref{fig:criticaledge} presents a plot of the current $J_E(X,t) \times 2t/[J v_{max}(\beta_r - \beta_l)]$, i.e., of $K^c(X)$ (solid line) as a function of the scaling variable $X$, which is compared with the asymptotic behavior for $X\to-\infty$ (dashed line).
%
%
%
%

One immediately notes that for $X>0$ the critical kernel $K^c(X)$ is qualitatively similar to the non critical one $K^A(X,X)$ which determines, up to constants, $J_E$ in Eq.~(\ref{eisler}) and they both decay exponentially upon increasing $X$, as can be readily checked from Eq.~(\ref{Airy+inf}). 
On the contrary, for $X<0$, the typical staircase structure of the Airy kernel $K^A(X,X)$ shown in Fig.~\ref{fig:heatedge} is absent in the critical case $K^c(X)$ reported in Fig.~\ref{fig:criticaledge}, due to the fact that the integration in Eq.~\eqref{eq:rel-kernels} smoothens it to the extent that it is no longer visible.
Accordingly, when $h$ is set to its critical value $h_c=1$, although the relevant scaling variable is $\propto (x-v_{max}t)/t^{1/3}$ as in the non-critical case of Eq.~(\ref{scalingvariables}) (this is due to the fact that Eq.~(\ref{stationaryphase}) still admits a unique solution $k_s=0$ at which the second derivative of the dispersion relation vanishes when $x\to v_{max}t$), the qualitative features of $J_E(X,t)$ change significantly as a consequence of the fact that the energy gap vanishes and a novel kernel $K^c(X)$ (related to the Airy kernel by Eq.~\eqref{eq:rel-kernels}) emerges. 
Note that  $K^c(X)$  is as "universal" as the Airy kernel, as it does not depend on the specific properties of the system under investigation, i.e., $J$, $v_{max}$, etc.~and, together with the scaling function which involves it, is essentially determined by $\varepsilon'(k_s)$ and $\varepsilon^{(3)}(k_s)$.

\section{Summary and perspectives}
\label{sec:comm}

In this work we have investigated the non-equilibrium dynamics induced by joining via a local interaction two independent transverse field quantum Ising chains, initially thermalized at two different temperatures. In Sec.~\ref{firstsection} the model and the quench protocol have been introduced: in particular it has been shown, 
following Ref.~\cite{de2013nonequilibrium}, how the restored translational invariance of the final Hamiltonian naturally determines a two-fold degeneracy of the single particle spectrum $\varepsilon (k)$, which allows the description of the excitations of the full chain in terms of  right- and left-moving quasi-particles. 
 
In terms of these excitations, in Sec.~\ref{secondsection}, the current of energy $J_E(x,t)$, of fermionic quasi-particles $J_N(x,t)$ and the corresponding densities $u(x,t)$ and $n(x,t)$, respectively,  have been exactly calculated in the space-time scaling limit in which, formally, both $x$ and $t$ are assumed to be large on the corresponding microscopic scale, with a fixed ratio $v=x/t$. In particular, one of the main result of the present work concerns the form of the profile of the energy and the quasi-particles currents and densities in Fig.~\ref{fig:scaling} as functions of $v$ and the fact that their qualitative behavior depends on the transverse field $h$ being critical or not, as shown in Figs.~\ref{fig:edge1}, \ref{fig:color}, \ref{fig:edge2}, \ref{fig:edge-JN}, and \ref{fig:edge-N}. All these figures show how these quantities propagate along the chain in the form of a front travelling with the characteristic velocity given in Eq.~(\ref{characteristicvel}). At any finite t
 ime $t$, the sites of the chain which are further away from the origin than  $v_{max}t$ are unperturbed and still retain their initial features. These asymptotically far regions play the role of thermal baths which allow the development, and eventually the persistence in the steady state, of a non-equilibrium dynamics. 
In this context, our results in Eqs.~\eqref{finalresult}, --- explicitly calculated in Eqs.~\eqref{rl}, \eqref{integration}, and \eqref{eq:def-G-mt} --- \eqref{finalresultenergy}, \eqref{finalparticle}, and \eqref{finalJn}, for $J_E$, $u$, $J_N$, and $n$, respectively, generalize the well-known picture of current-carrying steady states \cite{bernard2012energy,bernard2015non,doyon2012nonequilibrium} 
by describing the whole dynamics, both in space and time, and also how the stationary state is actually approached.  
In Sec.~\ref{thirdsection} we investigate the behavior of the two-point correlation function in Eq.~\eqref{starting3} and of the energy current $J_E(x,t)$ in Eq.~\eqref{operator2}, close to the edge of the front, i.e., for $x \simeq v_{max}t$ and beyond the space-time scaling limit discussed in Sec.~\ref{secondsection} summarized above. In particular we show that, as it occurs in similar cases investigated in the literature \cite{viti2015inhomogeneous,eisler2013full,allegra2016inhomogeneous,hunyadi2004dynamic}, these quantities acquire a "universal" behavior within a region of width $\Delta x \propto t^{-1/3}$ around the edge at $x \simeq v_{max}t$, conveniently expressed in terms of the scaling variable $X \propto (x-v_{max}t)/t^{1/3}$.
Since the two Ising chains are initially in a thermal state, the Fermi-Dirac statistics enters in the expression of these two physical observables, respectively in Eq.~(\ref{airyexpansion}) and Eq.~(\ref{eisler}). At the leading order, for long times on the microscopic scale, they coincide (up to multiplicative constants) with the kernel which characterizes the behaviour of the edge of either a system of free fermions initially prepared either in a factorized Fermi-sea ground state with different fillings on the two parts of the chain \cite{viti2015inhomogeneous} or in a domain-wall initial state \cite{eisler2013full,allegra2016inhomogeneous}. In order to investigate the effects of temperature on correlations beyond this simple rescaling, we also determined the first-order non-trivial correction for the two-point correlation function, which can be expressed in terms of the Airy kernel. 
As observed in Sec.~\ref{secondsection} for the profiles of various quantities as functions of $v = x/t$ in the space-time scaling limit, a change in the qualitative behavior of the current $J_E$ occurs as a function of $X$ as the transverse field takes its critical value $h_c=1$; in fact, the leading term in Eq.~\eqref{eisler} vanishes and  
therefore higher-order terms must be considered. Specifically, a scaling form emerges which involves a 
novel kernel $K^c(X)$, see Eq.~(\ref{kernelcritical}), which is actually the integral of the Airy kernel and in which the staircase structure characterising the latter basically disappears, see Figs.~\ref{fig:heatedge} and \ref{fig:criticaledge}.  
 
Among the possible extensions of the present work we plan to carry out the analysis of the scaled cumulant generating function \cite{bernard2012energy,touchette2009large} of the energy in the space-time scaling limit, in order to study how the whole statistics of this quantity and therefore its fluctuations changes upon approaching the edge of the propagating front and upon tuning $h$ to its critical value. So far, only fluctuations in the steady state have been studied \cite{de2013nonequilibrium}. Moreover, we will also explore the possibility to extend the general diagonalization procedure explained in Sec.~\ref{firstsection} to other one-dimensional quantum chains models, even not translationally invariant like that one studied in Ref.~\cite{chung2012thermalization}. A description of these models in terms of right/left moving quasi-particles is still lacking and it is in fact not yet clear which form these excitations may take for such systems and how this could influence transpor
 t properties and their statistics.

\begin{acknowledgements}
The authors would like to thank V.~Eisler, M.~Kormos, A.~De Luca and J.~Viti for useful discussions. 
\end{acknowledgements}


\newpage

\onecolumngrid

\numberwithin{equation}{section}
\titleformat{\section}{\large\bfseries}{\appendixname~\thesection .}{0.5em}{}

\begin{appendices}

\section{Exact solution of the transverse field Ising chain}
\label{appendix}

The Hamiltonian in Eq.~(\ref{Shultz}) can be written as a quadratic form in the fermionic operators $c_i$ and $c_i^\dagger$ as
\begin{equation}
H_r = \sum_{i,j} \left[ c_i^\dagger M_{ij} c_j + \frac{1}{2} \left(c_i^\dagger N_{ij} c_j^\dagger +h.c. \right)\right],
\end{equation}
where
\begin{equation}
M_{ij} = -\frac{J}{2}(\delta_{i+1,j} + \delta_{i,j+1}) -Jh\delta_{ij} \qquad \mbox{and} \qquad N_{ij} = -\frac{J}{2}(\delta_{i,j+1} - \delta_{i+1,j}). \label{matrices}
\end{equation}
An analogous expression holds for $H_l$. 
In order to diagonalize this type of Hamiltonians we perform the canonical transformation reported in Eq.~\eqref{phioperator} and then impose that the resulting Hamiltonian, once expressed in terms of $\phi_r(k)$, takes the diagonal form reported in Eq.~\eqref{diagonal}. This amounts at requiring that  
\be
\left[\phi_{r}, H_{r} \right] - \varepsilon(k) \phi_{r}(k) = 0,
\ee
which, implies the following conditions for the coefficients $\omega_{r}^q (k)$ and $\xi_{r}^q(k)$ in Eq.~\eqref{phioperator}:
\begin{align}
\varepsilon (k) \omega_{r}^q (k) = \sum_j \left( \omega_{r}^j (k) M_{jq} - \xi_{r}^j(k) N_{jq}\right), \label{eq:app-cond1}\\
\varepsilon (k) \xi_{r}^q (k) = \sum_j \left( \omega_{r}^j (k) N_{jq} - \xi_{r}^j(k) M_{jq}\right).  \label{eq:app-cond2}
\end{align}
At this point it is convenient to express $\omega_{r}^q (k)$ and $\xi_{r}^q(k)$ in terms of their sum and difference, i.e., as 
\begin{eqnarray}
\omega_r^q (k)  = \frac{A_r^q(k)+B_r^q(k)}{2} \qquad \mbox{and} \qquad 
\xi_r^q (k) = \frac{A_r^q(k)-B_r^q(k)}{2},
\label{AB}
\end{eqnarray}
which turn the conditions \eqref{eq:app-cond1} and \eqref{eq:app-cond2} into the following matrix form 
\begin{subequations}
\label{eq:app-MN}
\begin{align}
A_{r} (k) (M-N) & =  \varepsilon(k) B_{r} (k), \\ 
B_{r} (k) (M+N) & =  \varepsilon(k) A_{r} (k),
\end{align}
\end{subequations}
where $A_{r} = \left(A_r^1,\; A_r^2,\; A_r^3\;... \; A_r^N \right)$ is the vector of the coefficients $\{A_r^q \}_{q=1,\ldots, N}$, and analogous for $B_r$. Written in terms of components, these equations become:
\be
\left \{
         \begin{array}{r}
          hJ A_r^{q} (k) -JA_r^{q+1} (k) = \varepsilon(k) B_r^{q} (k),  \\[1mm]
          -J B_r^{q} (k) +hJ B_r^{q+1} (k) = \varepsilon(k) A_r^{q} (k),
          \end{array}
          \right.
\ee
with $q=1,2, \ldots, N$ and boundary conditions $A_{r}^{N+1} (k) = 0 $ with $B_{r}^{0} (k) = 0$. 
The solution of this system of equation is \cite{lieb2004two}
\be
A_r^q (k) = a_k \sin (qk-f(k)) \quad \mbox{and} \quad B_r^q (k) = b_k \sin (qk), \label{discrete}
\ee
where $a_k$ and $b_k$ are (ortho-)normalization constants which are determined by requiring the normalization of the vectors, i.e., $\sum_{q=1}^{N} A_r^{q}(k) A_r^{q}(k')=\sum_{q=1}^{N} B_r^{q}(k) B_r^{q}(k')=\delta_{k,k'}$, while $k$ and $f(k)$ are given by Eqs.~\eqref{kvalues} and \eqref{angle}, respectively.

In the thermodynamic limit $N \rightarrow \infty$ the set of allowed values of $k$ becomes continuous and covers the interval $[0,\pi]$, while the functions in Eq.~\eqref{discrete} become 
\begin{eqnarray} 
A_r^q (k) = \sqrt{\frac{2}{\pi}} \sin(qk-f(k)) \quad \mbox{and} \quad B_r^q (k)=\sqrt{\frac{2}{\pi}} \sin(qk). \label{symmetry}
\end{eqnarray} 
The procedure outlined above carries over to the left chain in Eq.~(\ref{chains}); in particular one introduces 
\be 
\phi_l (k) = \sum_{q=-N+1}^{0} \left( \omega_l^q (k) c_q + \xi_l^q(k) c_q^\dagger \right)
\label{bog2}
\ee
and the matrices $M_{ij}$ and $N_{ij}$ analogous to  Eq.~(\ref{matrices}), but appropriate for the right chain, as well as the vectors $A_l$ and $B_l$ as discussed above for the right chain. 
Due to the relationship between the expressions of $H_r$ and $H_l$ in Eqs.~\eqref{chains}, 
the functions $\{ A_l^q, B_l^q\}$ turn out to be related to those of the right chain as
\be
A_l^q (k) = B_r^{1-q} (k) \quad \mbox{and}\quad B_l^q (k) = A_r^{1-q} (k), 
\ee
which follows from the fact that the left chain has boundary conditions dual to those of the right chain, i.e., $A_{l}^{1} (k) = 0 $ and $B_{l}^{-N} (k) = 0$. \\   

After joining the two chains, the post-quench Hamiltonian in Eq.~(\ref{postquenchH-0}) can be written as
\begin{equation}
H = -\frac{J}{2} \left( \sum_{q=-N+1}^{N-1} \sigma_q^x \sigma_{q+1}^x + h\sum_{q=-N+1}^{N} \sigma_q^z \right),  \label{postquenchH}
\end{equation}  
which has the same expression as $H_r$ in Eq.~(\ref{chains}) but with $N \rightarrow 2N$ and $q \rightarrow q+N$ and  
therefore it can be diagonalized as explained above. More precisely, by expanding Eq.~(\ref{kvalues}) for $k_n$  in the thermodynamic limit $N \rightarrow +\infty$, after the replacement $N \rightarrow 2N$, one finds
\begin{equation}
k_n = \frac{\pi}{2N} -\frac{\pi n}{4N^2} + \frac{1}{2N}\mbox{arctan}\left(\frac{\mbox{sin}(\frac{\pi n}{2N})}{\mbox{cos}(\frac{\pi n}{2N})-h}\right) +O\left(\frac{1}{N^3}\right). 
\label{approximate}
\end{equation}
Accordingly, $k_n = \pi n/2N \rightarrow k \in (0,\pi)$ at the lowest order in $1/N$. 
By using Eq.~(\ref{approximate}) into Eq.~(\ref{discrete}) and after shifting the lattice index $q \rightarrow q+N$, one readily observes that the analogous of Eqs.~\eqref{eq:app-MN} for $H$ are satisfied by two sets of functions $A_1^q(k),\,  B_1^q(k)$ and $A_2^q(k),\, B_2^q(k)$ according to the parity of the index $n$ labelling the discrete momenta before taking the thermodynamic limit: 
\begin{eqnarray}
A_1^q (k) = \sqrt{\frac{1}{\pi}} \sin\left(qk-\frac{f(k)+k}{2}\right), \qquad B_1^q (k)=\sqrt{\frac{1}{\pi}} \sin\left(qk + \frac{f(k)-k}{2}\right), \\
A_2^q (k) = \sqrt{\frac{1}{\pi}} \cos\left(qk-\frac{f(k)+k}{2}\right), \qquad B_2^q (k)=\sqrt{\frac{1}{\pi}} \cos\left(qk + \frac{f(k)-k}{2}\right).
\label{AB2}
\end{eqnarray}
As a consequence, we can construct, via a Bogoliubov transformation similar to the one introduced in Eqs.~(\ref{phioperator}) and (\ref{bog2}), the two following operators:
\begin{eqnarray}
\Psi_1 (k) = \sum_{q=-\infty}^{+\infty} \left[\omega_1^q (k) c_q + \xi_1^q (k) c_q^\dagger \right], \nonumber \\
\Psi_2 (k) = \sum_{q=-\infty}^{+\infty} \left[\omega_2^q (k) c_q + \xi_2^q (k) c_q^\dagger \right],
\label{postquench}
\end{eqnarray}
in terms of which $H$ is diagonal. The functions $\omega_{1,2}^q (k)$ and $\xi_{1,2}^q (k)$ are determined in the same way as in Eq.~(\ref{AB}). 
Due to translational symmetry of the total chain  for $N \rightarrow \infty$, it is convenient the look for a linear combination $\Psi_{R,L} (k)$ of the operators $\Psi_{1,2}(k)$, which transforms according to Eq.~\eqref{transinvariance}
under spatial translations, which turns out to be given by
\begin{equation}
\begin{pmatrix} \Psi_R (k) \\ \Psi_L (k) \end{pmatrix}  =\frac{e^{-i \frac{f(k)-k}{2}}}{\sqrt{2}} \begin{pmatrix} -i & 1 \\ i & 1 \end{pmatrix}  \begin{pmatrix} \Psi_1 (k) \\ \Psi_2 (k) \end{pmatrix} .
\label{RightLeftmovers}
\end{equation}
These two operators can be interpreted as fermionic quasi-particles excitations delocalized along the chain.
In particular, from the explicit expression of the functions in Eq.~(\ref{AB2}), one eventually finds 
\begin{eqnarray}
\Psi_{R,L} (k) & = & \sum_{q=-\infty}^{+\infty} \left[ c_q \; \omega_{R,L}^q (k) + c_q^\dagger \;  \xi_{R,L}^q (k) \right], \nonumber \\
A_{R}^q (k) & = & \sqrt{\frac{1}{2 \pi}} e^{i \left(-qk +k  \right)}, \qquad B_{R}^q (k) = \sqrt{\frac{1}{2 \pi}} e^{i \left[-qk+k-f(k)\right]}, \nonumber \\  
A_{L}^q (k) & = & \sqrt{\frac{1}{2 \pi}} e^{i \left[qk-f(k)\right]}, \qquad B_{L}^q (k) = \sqrt{\frac{1}{2 \pi}} e^{i qk}.   \label{opposite}
\end{eqnarray}

For future reference, matrix elements between the pre- and post-quench single-particle fermionic states are calculated here, since they play a fundamental role in determining the dynamics of transport properties discussed further below. In order to do this, one can invert Eq.~(\ref{phioperator}) and express the local fermionic operators $c_q$ with $q=1,\ldots, N$ in terms of $\phi_r(k)$, i.e., of the pre-quench operators:
\begin{equation}
c_q = \int_{0}^\pi dk \, \left[\phi_r (k) \omega_r^q (k) + \phi_r^\dagger (k) \xi_r^q (k)\right], 
\label{insert}
\end{equation} 
where the coefficients of the linear combination of $\phi_r (k)$  and $\phi_r^\dagger (k)$ are determined such that to preserve the canonical fermionic anti-commutation relation $\{c_q, c^\dagger_{q'}\} = \delta_{q,q'}$; in particular, this implies the following completeness relation 
\begin{equation}
\int_{0}^{\pi} dk \, \left[\omega_r^q (k) \omega_r^{q'}(k) + \xi_r^q (k) \xi_r^{q'} (k)\right] = \delta_{q,q'},
\end{equation}  
for  $\omega_r^q (k)$ and $\xi_r^q (k)$.
Substituting Eq.~(\ref{insert}) into Eqs.~(\ref{postquench}) and doing the same with $r \rightarrow l$ one can express the right- and left-moving fermions in terms of the pre-quench operators:
\begin{equation}
\begin{pmatrix} \Psi_R (k) \\ \Psi_L (k)  \end{pmatrix} = \int_0^\pi dk' \left[ m_1(k,k') \begin{pmatrix} \phi_r (k') \\ \phi_l (k')  \end{pmatrix}  + m_2(k,k')\begin{pmatrix}  \phi_r^\dagger (k') \\ \phi_l^\dagger (k') \end{pmatrix}  \right] ,
\label{changebasis}
\end{equation}
where $m_1(k,k')$ and $m_2(k,k')$ are $2 \times 2$ matrices containing the coefficients of the superposition of the pre-quench operators. 
In particular, they take the form
\be
m_i(k,k')= \begin{pmatrix} m_{i,Rr} (k,k') & m_{i,Rl} (k,k') \\ m_{i,Lr} (k,k') & m_{i,Ll} (k,k') \end{pmatrix}  \quad \mbox{with} \quad i=\{1,2\}. 
\ee

Each matrix element in $m_{1,2}(k,k')$ can be expressed as a series involving the functions in Eqs.~(\ref{AB2}) and (\ref{AB}). In particular $m_{1,Rr}$ and $m_{1,Rl}$ turn out to be the relevant ones for the calculation of the heat current presented in Sec.~\ref{secondsection} and they are given by
\begin{subequations}
\label{matrixcoeff}
\begin{align}
m_{1,Rr}(k,k')= \sum_{q=1}^{+\infty} \left[ \omega_R^q (k) \omega_r^q (k') + \xi_R^q (k) \xi_r^q (k')\right], \\
m_{1,Rl}(k,k')= \sum_{q=0}^{-\infty} \left[\omega_R^q (k) \omega_l^q (k') + \xi_R^q (k) \xi_l^q (k')\right]. 
\end{align}
\end{subequations}
As anticipated, these coefficients can indeed be interpreted as matrix elements: in fact, given the pre-quench single particle state $ \vert k' \rangle_{r,l} \equiv \phi_{r,l}^\dagger (k') \vert 0 \rangle$ and the post-quench one $ \vert k \rangle_R = \Psi_R^\dagger (k) \vert 0 \rangle$, where $\vert 0 \rangle$ is the ground state of the complete chain, from Eq.~(\ref{changebasis}) the various coefficients are recognized to correspond to the following scalar product, where $\alpha \in \{ r, l\}$,
\be
{}_R\langle k \vert k' \rangle_{\alpha} =  \langle 0 \vert \Psi_R (k)\phi_\alpha^\dagger (k') \vert 0 \rangle 
=  \int_0^\pi dk'' \langle 0 \vert m_{1,R\alpha}(k,k'') \phi_\alpha (k'')\phi_\alpha^\dagger (k') \vert 0 \rangle
=   m_{1,R\alpha} (k,k'). 
\ee
\section{Space-time scaling limit} 
\label{appendix2}

In this appendix we report the details of the calculations leading to the expression of the energy current $J_E(x,t)$ in Eq.~(\ref{finalresult}). One starts by substituting Eq.~(\ref{fourier}) into Eq.~(\ref{localcurrent}) and by evolving in time  the corresponding operator under the unitary dynamics prescribed by $H$:
\begin{align}
j_E (x,t) = & \; \;  \frac{ihJ^2}{2} \left( c_{x+1}^\dagger c_{x} - c_x^\dagger c_{x+1} \right) \nonumber \\ 
          = & \; \;  \frac{ihJ^2}{2} \int_{-\pi}^{\pi} dk \int_{-\pi}^{\pi} dk' \left( e^{-ik} -e^{ik'}\right) e^{i[\varepsilon(k)-\varepsilon(k')]t} \Psi_R^\dagger (k) \Psi_R (k') \left\{\omega_R^x(k)\left[\omega_R^x(k')\right]^\ast + \xi_R^x(k)\left[\xi_R^x(k')\right]^\ast\right\} + \nonumber \\
          & \; \; \frac{ihJ^2}{2} \int_{-\pi}^{\pi} dk \int_{-\pi}^{\pi} dk' \left( e^{-ik} -e^{-ik'}\right) e^{i[\varepsilon(k)+\varepsilon(k')]t} \Psi_R^\dagger (k) \Psi_R^\dagger (k') \omega_R^x(k) \xi_R^x(k') + \nonumber \\ 
          & \; \; \frac{ihJ^2}{2} \int_{-\pi}^{\pi} dk \int_{-\pi}^{\pi} dk' \left( e^{ik} -e^{ik'}\right) e^{-i[\varepsilon(k)+\varepsilon(k')]t} \Psi_R (k) \Psi_R(k') \left[\omega_R^x(k')\right]^\ast \xi_R^x (k), 
\label{operator} 
\end{align}
where we have used the fermionic anticommutation relation $\{ \Psi_R(k), \Psi_R^\dagger(k') \} = \delta(k-k')$. As we shall explain later in this appendix the last two terms of Eq.~(\ref{operator}) are vanishing in the limit $k=k'$ and therefore they will not contribute in the space-time scaling limit that will be introduced below. As a consequence, we will henceforth mainly consider the first term of the sum in Eq.~(\ref{operator}). The integral is calculated over the square domain $[-\pi,\pi] \times [-\pi,\pi]$, which for brevity, will not be indicated further below. \\
The trace in Eq.~(\ref{trace}) can be calculated based on the knowledge of ${\rm Tr}[\Psi_R^\dagger (k) \Psi_R (k') \rho_0]$,${\rm Tr}[\Psi_R^\dagger (k) \Psi_R^\dagger (k') \rho_0]$ and ${\rm Tr}[\Psi_R (k) \Psi_R (k') \rho_0]$. In turn, considering the first contribution, one exploits the change of basis in Eq.~(\ref{changebasis}) by writing:
\begin{equation}
\Psi_R^\dagger (k) \Psi_R (k') = \int_0^{\pi} dk_1 dk_2 \left[m_{1,Rr}^\ast (k,k_1) m_{1,Rr}(k',k_2) \Phi_r^\dagger (k_1) \Phi_r(k_2) + m_{1,Rl}^\ast (k,k_1) m_{1,Rl}(k',k_2) \Phi_l^\dagger (k_1) \Phi_l(k_2) +...\right],  
\end{equation}
where $"..."$ denotes terms which are regular within the integration domain and therefore, as explained further below, do not contribute to the integral. Taking the trace of this expression as in Eq.~(\ref{trace}) and remembering that (since from Eq.~(\ref{HRcont}) one has that the operators $\Phi_{r,l}$ represent the excitations of a free fermion system) 
\begin{equation}
\mbox{Tr}[\rho_0 \Phi_\alpha^\dagger (k) \Phi_\gamma (k')] = \delta_{\alpha \gamma} \delta (k-k') f_\alpha(k) = \delta_{\alpha \gamma} \delta (k-k') \frac{1}{e^{\beta_{\alpha}\epsilon(k)}+1}, \qquad \alpha \in \{r,l\} \label{traceapp}
\end{equation}
one gets \begin{equation}
\mbox{Tr}[\rho_0 \Psi_R^\dagger (k) \Psi_R (k')] = I(k,k') = I_r(k,k') + I_l(k,k'), \label{integraldef}  \\
\end{equation} 
where we defined 
\begin{equation}
I_{\alpha}(k,k')= \int_{0}^{\pi} dk_1 m_{1,R{\alpha}}^\ast (k,k_1) m_{1,R{\alpha}}(k',k_1) f_{\alpha}(k_1). \label{integral}
\end{equation}
Within the space-time scaling limit introduced in Sec.~\ref{secondsection} to calculate Eq.~(\ref{operator2}) we are led to consider the points where the phase $\varphi_{x,t} (k,k') = [\varepsilon (k) - \varepsilon (k')]t + x(k'-k)$, appearing in the expression of $J_E(x,t)$ which follows from calculating the trace of Eq.~(\ref{operator}) according to Eq.~(\ref{trace}), is stationary, taking into account the explicit expression of $\omega_R^x (k)$ and $\xi_R^x (k)$ from Eq.~(\ref{opposite}),
one obtains: 
\begin{equation}
\left \{
         \begin{array}{ll}
          \displaystyle {v_g(k) t -x = 0},  \\ 
          \\
         \displaystyle {-v_g(k') t+x = 0},
          \end{array}
          \right. \label{system}
\end{equation}
where we introduced the group velocity $v_g(k)$ of the excitations as already done in Eq.~(\ref{finalresult}). Depending on the value of the ratio $v \equiv x/t$, each of the two equations in Eq.~(\ref{system}) has either no or two solutions $k_s^+(v)$ and $k_s^-(v)$ (as shown in the main text in Fig.~\ref{fig:timedensity}). In the latter case we have four possible stationary points for the pair $(k,k')$, i.e., ($k_s^+$,$k_s^+$), ($k_s^-$,$k_s^-$), ($k_s^+$,$k_s^-$), ($k_s^-$,$k_s^+$). Focusing now on Eq.~(\ref{integral}), one needs to specify further the structure of the matrix elements $m_1,Rr(k,k')$ and $m_1,Rl(k,k')$ involved. From Eq.~(\ref{matrixcoeff}) calculating the series one finds
\begin{eqnarray}
m_{1,Rr} (k,k') & = & \frac{1}{4 \pi i} \left[\frac{e^{i[k-f(k')]}}{1-e^{i(k'-k+i \delta)}} - \frac{e^{i[k+f(k')]}}{1-e^{-i(k+k'-i \delta)}}+ \frac{e^{i[k-f(k)]}}{1-e^{i(k'-k+i \delta)}} - \frac{e^{i[k-f(k)]}}{1-e^{-i(k+k'-i \delta)}}\right] , \nonumber \\
m_{1,Rl} (k,k') & = & \frac{1}{4 \pi i} \left[\frac{e^{-i[f(k)+f(k')]}}{1-e^{i(k'+k+i \delta)}} - \frac{e^{i[f(k')-f(k)]}}{1-e^{i(k-k'+i \delta)}}+\frac{1}{1-e^{i(k'+k+i \delta)}}- \frac{1}{1-e^{i(k-k'+i \delta)}} \right], 
\label{matrixelements}
\end{eqnarray}
here $\delta>0$ is an infinitesimal positive constant needed to ensure convergence of the series in Eq.~(\ref{matrixcoeff}) eventually set to zero. Using the fact that $m_{1,R{\alpha}} (k,k')= -m_{1,R{\alpha}} (k,-k')$ and $f_{\alpha} (k) = f_{\alpha} (-k)$, with $\alpha \in \{l,r\}$, one finds from Eq.~(\ref{integral}) that:
\begin{eqnarray}
I_{\alpha}(k,k')= & = & \frac{1}{2} \int_{-\pi}^{\pi} dk_1 m_{1,R{\alpha}}^\ast (k,k_1) m_{1,R{\alpha}}(k',k_1) f_{\alpha}(k_1) = \nonumber \\ 
          & = & \frac{1}{2} \oint_{C_1} dz \frac{m_{1,R{\alpha}}^\ast (k,-i\mbox{ln}(z)) m_{1,R{\alpha}}(k',-i\mbox{ln}(z)) f_{\alpha}(-i\mbox{ln}z)}{iz},
\label{complexintegral}
\end{eqnarray}
where the original integral has been extended to the complex plane, along the circumference $C_1$ centred in the origin and with unitary radius, via the change of variable $z=e^{ik_1}$. Since the integration path is a closed contour, $I (k,k')$ is determined by the singularities of the integrand which are located inside $C_1$. Applying the residue theorem to Eq.~(\ref{complexintegral}) with the expressions in Eq.~(\ref{matrixelements}), the final form of $I(k,k')$ from Eq.~(\ref{integraldef}) is:
\begin{equation}
I(k,k') = \frac{1}{4 \pi i} \left( \frac{f_{\beta_l}(k)+f_{\beta_l}(k')}{k-k'-2i \delta} - \frac{f_{\beta_r}(k)+f_{\beta_r}(k')}{k-k'+2i \delta} \right) + \quad \mbox{regular} \quad \mbox{terms} \quad \mbox{as} \quad k \rightarrow k'. \label{integralapp}
\end{equation}
In the space-time scaling limit introduced to calculate Eq.~(\ref{operator}) one expects the double integral in $k$ and $k'$ to be dominated by the pairs $(k,k')$ solution of Eq.~(\ref{system}) where the phase $\varphi(k,k')$ is stationary as well as possible singularities in $I(k,k')$. By inspection of Eq.~(\ref{integralapp}) one notices that the only saddle points where $I(k,k')$ is stationary are the couples ($k_s^+$,$k_s^+$), ($k_s^-$,$k_s^-$), therefore one concludes that the double integral in $k$ and $k'$ of Eq.~(\ref{operator}) is dominated by the region where $k \simeq k'$. This is also the reason why we omitted in Eq.~(\ref{integralapp}) terms which are not singular as $k \rightarrow k'$ since they do not contribute in the space-time scaling limit, as just explained. For the same reason, in the space-time scaling limit, we can neglect the second and the third term of Eq.~(\ref{operator}) since they vanish in the limit $k \rightarrow k'$ as anticipated at the beginning of th
 e section.

More specifically, considering the term $I_2 (k,k') \equiv \mbox{Tr}[\rho_0 \Psi_R^\dagger (k) \Psi_R^\dagger (k')]$, it can be evaluated following the same steps considered for $I(k,k')$
\begin{equation}
\Psi_R^\dagger (k) \Psi_R^\dagger (k') = \int_0^{\pi} dk_1 dk_2 \left[m_{1,Rr}^\ast (k,k_1) m_{2,Rr}^\ast(k',k_2) \Phi_r^\dagger (k_1) \Phi_r(k_2) + m_{1,Rl}^\ast (k,k_1) m_{2,Rl}^\ast(k',k_2) \Phi_l^\dagger (k_1) \Phi_l(k_2) +...\right] \label{secondexpansion}
\end{equation}  
where the new matrix elements $m_{2,Rr}(k,k')$,$m_{2,Rl}(k,k')$ are given by the series
\begin{eqnarray}
m_{2,Rr}(k,k')&=& \sum_{q=1}^{+\infty} \left[ \omega_R^q (k) \xi_r^q (k') + \xi_R^q (k) \omega_r^q (k')\right], \nonumber \\
m_{2,Rl}(k,k')&=& \sum_{q=0}^{-\infty} \left[\omega_R^q (k) \xi_l^q (k') + \xi_R^q (k) \omega_l^q (k')\right], \label{secondmatrixelement}
\end{eqnarray}
that turns out to sum to
\begin{eqnarray}
m_{2,Rr}(k,k')&=& \frac{1}{4 \pi i} \left[\frac{e^{i[k-f(k')]}}{1-e^{i(k'-k+i \delta)}} - \frac{e^{i[k+f(k')]}}{1-e^{-i(k+k'-i \delta)}}- \frac{e^{i[k-f(k)]}}{1-e^{i(k'-k+i \delta)}} + \frac{e^{i[k-f(k)]}}{1-e^{-i(k+k'-i \delta)}}\right]           , \nonumber \\
m_{2,Rl}(k,k')&=& \frac{1}{4 \pi i} \left[\frac{e^{-i[f(k)+f(k')]}}{1-e^{i(k'+k+i \delta)}} - \frac{e^{i[f(k')-f(k)]}}{1-e^{i(k-k'+i \delta)}}-\frac{1}{1-e^{i(k'+k+i \delta)}}+\frac{1}{1-e^{i(k-k'+i \delta)}} \right].             \label{secondmatrixelementssummed}
\end{eqnarray}
From Eq.~(\ref{secondexpansion}) $I_2 (k,k') \equiv \mbox{Tr}[\rho_0 \Psi_R^\dagger (k) \Psi_R^\dagger (k')]= I_2^l(k,k')+I_2^r(k,k')$ can be eventually calculated:
\begin{eqnarray}
I_2^r(k,k') & = &  \frac{1}{8 \pi} \left[ e^{-i[k-f(k)]}\left( e^{-i[k'-f(k)]}\frac{f_{\beta_r}(k)}{i(k-k')} - e^{-i[k'-f(k')]}\frac{f_{\beta_r}(k)}{i(k-k')}\right)\right],   \nonumber \\
I_2^l(k,k') & = & \frac{1}{8 \pi} \left[ e^{i[f(k')-f(k)]} \frac{f_{\beta_l}(k)}{i(k-k')} - \frac{f_{\beta_l}(k)}{i(k-k')}\right]. \label{integralapp2}
\end{eqnarray}
Once the trace over $\rho_0$ in Eq.~(\ref{traceapp}) has been taken, it is useful to define $Q = k-k'$ and $K=(k+k')/2$ and exploit the fact that $Q$ is infinitesimal in order to Taylor expand the dependence of the integral around $Q=0$, keeping only the leading order. According to this prescription one therefore notices that both $I_2^r(k,k')$ and $I_2^l(k,k')$, once they are multiplied by the factor $\, e^{-ik}-e^{-ik'} \,$ in Eq.~(\ref{operator}), vanish in the space-time scaling limit. The very same reasoning, that here is not reported again for brevity, can be also applied to the term determined by $\mbox{Tr}[\rho_0 \Psi_R (k) \Psi_R(k')]$.\\ 
Concentrating as a consequence on the first term of Eq.~(\ref{operator}), in particular, with reference to the phase $\varepsilon(k) - \varepsilon(k')$, one has: 
\begin{equation}
\varepsilon(k) - \varepsilon(k') = \varepsilon\left(K+\frac{Q}{2}\right) - \varepsilon\left(K -\frac{Q}{2}\right) = Qv_g (K) + O(Q^2).
\end{equation}
In terms of the new variables, the heat current takes the form (see Eq.~(\ref{operator}) and (\ref{trace})):
\begin{equation}
J_E (x,t) = \frac{hJ^2}{2 \pi} \int_{-\pi}^{\pi} dK \int_{-\infty}^{+\infty} dQ \left(\sin(K)\right) \frac{e^{iQ[v_g(K)t -x]}}{2 \pi i}\left(\frac{f_{\beta_l}(K)}{Q-2i \delta} -\frac{f_{\beta_r}(K)}{Q+2i \delta}\right).
\end{equation}
Recalling the integral representation of the Heaviside step function $\Theta(x)= \lim_{\delta \rightarrow 0^+} \int_{-\infty}^{\infty} dy \frac{1}{2 \pi i} \frac{e^{ixy}}{y-i \delta} $, we eventually find:
\begin{equation}
J_E (x,t) = \int_{-\pi}^{\pi} \frac{dk}{2 \pi} \varepsilon(k) v_g(k) \left[f_{\beta_l}(k)\Theta(v_g(k)t-x) - f_{\beta_r}(k)\Theta(x+v_g(k)t) \right], \label{finalresultapp}
\end{equation} 
which represents the exact profile of the energy current in time and space within the space-time scaling regime.

The integral over $k$ above can be restricted to the domain $[0,\pi]$ by using the fact that $\varepsilon(-k) = \varepsilon(k)$ and therefore $v_g(-k) = - v_g(k)$. The resulting expression clearly shows that $J_E(-x,t) = J_E(x,t)$ and therefore  $x$  in the r.h.s.~of Eq.~\eqref{finalresultapp} can be replaced by $|x|$. After introducing the convenient  scaling variable $v = x/t$ and using the fact that $\Theta(v-v_g(k))=1-\Theta(v_g(k)-v)$, one has
\be
J_E(v)=  \int_{0}^{\pi} \frac{dk}{2 \pi} \varepsilon(k) v_g(k) \left[f_{\beta_l}(k)-f_{\beta_r}(k)\right]\Theta(v_g(k)-|v|)= \Theta(v_{max}-|v|)\int_{k_s^-(v)}^{k_s^+(v)} \frac{dk}{2 \pi} \varepsilon(k) v_g(k) \left[f_{\beta_l}(k)-f_{\beta_r}(k)\right]
\label{finalresultapp-2}
\ee  
where $k^\pm_s(v)$ are the solutions of Eq.~\eqref{stationaryphase}, as shown in Fig.~\ref{fig:timedensity}. 
For later convenience, let us consider a generalization --- henceforth denoted by $J_a(v)$ --- of the previous expression in which $\varepsilon(k)$ in the integrand is replaced by $\varepsilon^a(k)$, with $a=1$ or 0. In fact, while $J_1 = J_E$, it turns out that $J_0 = J_N$, i.e., it corresponds to the particle current discussed in Sec.~\ref{subsec:corr-dens}, see Eq.~\eqref{finalJn}. Since $v_g(k)=\varepsilon'(k)$ one can perform the change of variable $k \mapsto \varepsilon(k)$ ending up with
\be
J_a(v)=\Theta(v_{max}-|v|)\int_{\varepsilon^-(v)}^{\varepsilon^+(v)}\frac{d\varepsilon}{2 \pi} \varepsilon^a\left[f_{\beta_l}(\varepsilon)-f_{\beta_r}(\varepsilon)\right], \quad \mbox{with} \quad \varepsilon^\pm(v) \equiv \varepsilon(k^\pm_s(v)),
\label{eqapp:Ja}
\ee
which after some rescaling can be written as
\be
J_a(v)= \Theta(v_{max}-|v|)\left[ {\cal J}_a(\beta_l,v)  - {\cal J}_a(\beta_r,v)\right],
\label{quasiexact-1}
\ee
with
\be
{\cal J}_a(\beta,v) = \frac{G_a(\beta \varepsilon^-(v))-G_a(\beta \varepsilon^+(v))}{2\pi \beta^{a+1}} ,
\label{quasiexact-2}
\ee
where we introduced
\be
G_a(x) \equiv \int_x^{+\infty} d\xi \, \frac{\xi^a}{e^\xi+1},
\label{eq:def-G-integr}
\ee
which takes the form reported in Eq.~\eqref{eq:def-G-mt} for $a=1$ ad in Eq.~\eqref{eq:G2} for $a=0$. In order to make this expression more explicit, one still need to determine $\varepsilon^\pm(v)$ defined in Eq.~\eqref{eqapp:Ja}, which is readily done by observing that the square of Eq.~\eqref{stationaryphase} is a second order equation for $\mbox{cos}(k^\pm_s(v))$, the solutions of which are
\be
\cos(k^\pm_s(v)) = \frac{v^2}{J^2 h} \mp \sqrt{\left (1-\frac{v^2}{J^2 h^2}\right)\left(1-\frac{v^2}{J^2}\right)}.
\label{criticalks}
\ee 
Replacing this expression 
into the dispersion relation Eq.~\eqref{spectrum}, one obtains 
\be
\varepsilon^\pm (v) =  \varepsilon_>(v) \pm \varepsilon_<(v),
\label{spectrumsolved}
\ee
where (see also Eq.~\eqref{characteristicvel})
\be
\varepsilon_>(v) = \sqrt{[J \max(1,h)]^2 -v^2} \quad \mbox{and} \quad \varepsilon_<(v) = \sqrt{ [J \min(1,h)]^2 -v^2}  = \sqrt{ v_{max}^2 -v^2}.
\ee
Inserting Eq.~\eqref{spectrumsolved} into Eq.~\eqref{quasiexact-1} (see also Eq.~\eqref{quasiexact-2}) for $a=1$ the expressions in Eqs.~(\ref{rl}) and (\ref{integration}) are eventually recovered. 
In order to determine the behaviour of $J_a(v)$ as $v$ approaches the edges at $\pm v_{max}$, note that, correspondingly, $\varepsilon_<(v)$ approaches zero $\propto \sqrt{|v-v_{max}|}$ while $\varepsilon_>(v) = \varepsilon_>(v_{max}) + {\cal O}(|v-v_{max}|)$. Accordingly, the leading term of the current $J_a$ in Eq.~\eqref{quasiexact-1}, upon approaching the edge $|v| \lesssim v_{max}$, can be  obtained from the expansion 
\be
G_a(\beta [\varepsilon_>(v)\pm \varepsilon_<(v)]) 
= G_a(\beta \varepsilon_>(v_{max})) \pm \beta \varepsilon_<(v) G'_a(\beta \varepsilon_>(v_{max})) + {\mathcal O}((v_{max}-|v|)^{3/2}), 
\label{eq:expG-app}
\ee
with  $G'_a(x) = -x^a/(e^x+1)$ [see Eq.~\eqref{eq:def-G-integr}]
and therefore $J_a$, according to Eqs.~\eqref{quasiexact-1} and \eqref{quasiexact-2}, can be expressed as 
\be
J_a(v) = C_a \left( v_{max}^2-v^2 \right)^{1/2} + {\cal O}((v_{max}-|v|)^{3/2}),
\label{semicircular-app}
\ee
where 
\be
C_a=\frac{G'_a(\beta_r \varepsilon_>(v_{max}))}{\pi \beta_r^a} - \frac{G'_a(\beta_l \varepsilon_>(v_{max}))}{\pi \beta_l^a} =  \frac{J^a |1-h^2|^{a/2}}{\pi} \left[ \frac{1}{e^{\beta_l J \sqrt{|1-h^2|}}+1} - \frac{1}{e^{\beta_r J \sqrt{|1-h^2|}}+1} \right] .
\label{eq:genexprC}
\ee
Note that the constant $C_a$ reported above vanishes if the field is turned to its critical value, i.e., if $h=1$. 
In this case, one has to consider in Eq.~\eqref{integration} with $\varepsilon_>(v) = \varepsilon_<(v)$ the next order in the expansion of $G_a(2\beta\varepsilon_<(v))$ for $\varepsilon_<(v) \to 0$. Taking into account that $G_1(x) = \pi^2/12 - x^2/4 + x^3/12 + {\cal O}(x^5)$ [see Eq.~\eqref{eq:def-G-integr}], one finds
\be
{\cal J}_1(\beta,v) = \frac{\varepsilon^2_<(v)}{2\pi} - \frac{\beta}{3 \pi} \varepsilon_<^3(v) + {\cal O}(\varepsilon_<^5(v)),
\ee
which, inserted into Eq.~\eqref{integration}, results into the different form of the edge behaviour reported in Eq.~\eqref{eq:edge-crit}. We also note that $C_1$ can be equivalently rewritten in terms of the single-particle energy spectrum and the Fermi-Dirac distributions as:
\be
C_1 = \frac{\varepsilon(k_s)}{\pi}(f_{\beta_l}(k_s)-f_{\beta_r}(k_s));
\label{C1constant}
\ee
the vanishing of $C_1$ --- which implies different behaviour of the energy current $J_E(v)$ as $v$ approaches the edge $|v| \to v_{max}^-$ for $h=1$ --- is therefore due to the fact the energy gap $\varepsilon(k_s)$ vanishes in this case.

Proceeding in exactly the same way, since $G_0(x) = \mbox{ln}2 - x/2 + x^2/8 -x^4/192 + {\cal O}(x^6)$, for the particle current one has 
\be
{\cal J}_0(\beta,v) = \frac{\varepsilon_<(v)}{2 \pi} - \frac{\beta}{4 \pi} \varepsilon^2_<(v) + {\cal O}(\varepsilon_<^4(v)),
\ee
which gives the critical edge behavior of Eq.~\eqref{eq:edgecritical2}, once inserted into Eq.~\eqref{eq:integration2}.

\section{Stationary phase calculation at the front edge} 
\label{appendix3}

\subsection{Correlation function}
\label{app:corr-edge}

In order to determine the two-point correlation function near the right edge of the front with $x, y \simeq v_{max}t$ (for the  front to the left with $x, y\simeq -v_{max}t$ the same procedure applies) we use a stationary-phase approximatiom starting from the expression
\be
\begin{split}
\langle \gamma_x^\dagger (t) \gamma_y (t) \rangle &= \int_{-\pi}^{\pi} \int_{-\pi}^{\pi} \frac{dk dk'}{2 \pi} e^{i[\phi_{x,t}(k)-\phi_{y,t}(k')]} I(k,k')  \\
                                        &= \int_{-\pi}^{\pi} \int_{-\pi}^{\pi} \frac{dk dk'}{2 \pi} e^{i[\phi_{x,t}(k)-\phi_{y,t}(k')]} \frac{1}{4 \pi i} \left[ \frac{f_{\beta_l}(k)+f_{\beta_l}(k')}{k-k'-2i \delta} + \frac{f_{\beta_r}(k)+f_{\beta_r}(k')}{k'-k-2i \delta} \right] \\
                                        &= {\langle \gamma_x^\dagger (t) \gamma_y (t) \rangle}_{l} + {\langle \gamma_x^\dagger (t) \gamma_y (t) \rangle}_{r}
\end{split}
\label{eq:corr-edge-corr}
\ee
which can be obtained following exactly the same steps which led to Eq.~(\ref{operator2}), with $I(k,k')$ given in Eq.~(\ref{integralapp}) and $\phi_{x,t}(k)= \varepsilon(k)t-k x$ with analogous definition for $\phi_{y,t}(k')$. Since, as explained in Sec.~\ref{thirdsection}, the stationary phase equation (\ref{stationaryphase}) admits only one solution $k_s$ for $x, y \simeq v_{max}t$, such that $\varepsilon'(k_s) = v_g(k_s) = v_{max}$ and $\varepsilon''(k_s) = v_g'(k_s) = 0$ (see Fig.~\ref{fig:timedensity}) at which both $\phi''_{x,t}(k)$ and  $\phi''_{y,t}(k)$ vanish, we need to expand them beyond the second order; actually we expand below up to the fourth order in order to account also the first correction beyond the leading behaviour, finding, e.g.,
\be
\phi_{x,t}(k)=  
\varepsilon(k_s)t-k_sx + (k-k_s)(v_{max}t-x)-\frac{v_{max}}{3!}(k-k_s)^3t + \frac{\varepsilon^{(4)}(k_s)}{4!} (k-k_s)^4 t + {\cal O}((k-k_s)^5),
\label{eq:app-exp-phi}
\ee
and analogous for $\phi_{y,t}(k')$, where we used the fact that $\varepsilon^{(3)}(k) = - \varepsilon'(k)[1+ 3 \varepsilon''(k)/\varepsilon(k)]$, which implies (for $h\neq 1$, the case $h=1$ is discussed further below) 
$\varepsilon^{(3)}(k_s) = - v_{max}$ and $\varepsilon^{(4)}(k_s) = 4 v_{max}^2/\varepsilon(k_s) = 4 v_{max}^2/(J\sqrt{|h^2-1|})$ (see Eq.~\eqref{spectrumsolved}). 
Accordingly, the first term in Eq.~\eqref{eq:corr-edge-corr} can be written as
\be
\begin{split}
\langle \gamma_x^\dagger (t) \gamma_y (t) \rangle_{l} = & e^{-i k_s(x-y)} \int_{-\infty}^{+\infty} \frac{d\widetilde{k}}{2 \pi} \int_{-\infty}^{+\infty} \frac{d\widetilde{q}}{2 \pi} \frac{f_{\beta_l}(\widetilde{k})+f_{\beta_l}(\widetilde{q})}{i(\widetilde{k}-\widetilde{q}-2i\delta)} \\
&\times \exp\left\{-i\left[\widetilde{k}  (x- v_{max}t) - \widetilde{q}(y-v_{max}t)+\widetilde{k}^3 \frac{v_{max}t}{3!} -\widetilde{k}^4 \frac{\varepsilon^{(4)}(k_s) t}{4!} -\widetilde{q}^3 \frac{v_{max}t}{3!} +\widetilde{q}^4 \frac{\varepsilon^{(4)}(k_s)t}{4!}\right]\right\} , \label{appendixcor}
\end{split}
\ee
with $\widetilde{k}=k-k_s$ and $\widetilde{q}=k'-k_s$.
Similarly, the Fermi-Dirac functions have also to be expanded around the stationary point
\begin{equation}
f_{\beta_l}(\widetilde{k}) = f_{\beta_l}(k_s) +f_{\beta_l}'(k_s)\widetilde{k} + \frac{f_{\beta_l}''(k_s)}{2}{\widetilde{k}}^2 + O(\widetilde{k}^3). \label{fermiexpansion}
\end{equation}
Since one expects the term proportional to $\widetilde{k}^3$ in the exponential to be the leading order, it is convenient to rescale the variables as $K=(v_{max}t/2)^{1/3} \widetilde{k}$, $Q=(v_{max}t/2)^{1/3} \widetilde{q}$ and introduce the scaled coordinates $X$ and $Y$ as in Eq.~\eqref{scalingvariables},  such that the exponentials in 
Eq.~(\ref{appendixcor}) can be written as:
\begin{align}
&&\exp\left(-iKX -\frac{iK^3}{3!}+\frac{i}{4!} \frac{\varepsilon^{(4)}(k_s)K^4\,2^{4/3}}{{v_{max}}^{4/3}t^{1/3}}\right) = \mbox{exp}\left(-iKX -\frac{iK^3}{3!}\right)\left(1+\frac{i}{4!} \frac{\varepsilon^{(4)}(k_s)K^4 \; 2^{4/3}}{{v_{max}}^{4/3} t^{1/3}}\right)+ {\cal O}\left((v_{max}t)^{-2/3}\right), \label{exponential1}\\
&&\exp\left(+iQY+\frac{iQ^3}{3!}-\frac{i}{4!}\frac{\varepsilon^{(4)}(k_s)Q^4 \; 2^{4/3}}{{v_{max}}^{4/3}t^{1/3}}\right) = \mbox{exp}\left(+iQY+\frac{iQ^3}{3!}\right)\left(1-\frac{i}{4!} \frac{\varepsilon^{(4)}(k_s)Q^4 \; 2^{4/3}}{{v_{max}}^{4/3} t^{1/3}}\right)+ {\cal O}\left((v_{max}t)^{-2/3}\right). \label{exponential2}
\end{align}
Inserting Eqs.~(\ref{fermiexpansion}), (\ref{exponential1}), and (\ref{exponential2}) into Eq.~(\ref{appendixcor}) and keeping  terms up to order $t^{-2/3}$ we end up with the result Eq.~(\ref{airyexpansion}) of the main text with   
\begin{equation}
K^{1}_l(X,Y) = i\left[\frac{f_{\beta_l}'(k_s)}{2 f_{\beta_r}(k_s)} \left(\frac{\partial K^{A}(X,Y)}{\partial X}-\frac{\partial K^{A}(X,Y)}{\partial Y}\right) + \frac{\varepsilon^{(4)}(k_s)}{12} \left(\frac{\partial^4 K^{A}(X,Y)}{\partial X^4}-\frac{\partial^4 K^{A}(X,Y)}{\partial Y^4}\right)\right], \label{kernelresult}
\end{equation} 
where we have used the integral representation of the Airy kernel
\be
K^{A}(X,Y)= \int_{-\infty}^{+\infty}\frac{dK}{2\pi}\int_{-\infty}^{+\infty}\frac{dQ}{2\pi}\frac{e^{-iKX-iK^3/3+iQY+iQ^3/3}}{i(K-Q-i\delta)},
\label{eq:def-KA-int}
\ee
which coincides with Eq.~\eqref{AiryKernel} once we observe that $-(\partial_X + \partial_Y)K^A(X,Y) = \mbox{Ai}(X)\mbox{Ai}(Y)$ with the usual integral representation of the Airy function:
\begin{equation}
\mbox{Ai}(X)= \int_{-\infty}^{+\infty} \frac{dK}{2 \pi} e^{iKX+iK^3/3}.
\label{eq:def-Airy}
\end{equation}
For completeness we report here the well-known asymptotic behaviours of the Airy function that have been used in the main text:
\begin{align}
&\mbox{Ai}(X\to +\infty)\simeq \frac{1}{2 \sqrt{\pi} X^{1/4}} \exp\left\{-\frac{2}{3} X^{3/2}\right\}, \label{Airy+inf}\\
&\mbox{Ai}(X\to-\infty) \simeq \frac{1}{\sqrt{\pi} |X|^{1/4}} \mbox{cos}\left(-\frac{2}{3}|X|^{3/2} + \frac{\pi}{4}\right)
\label{Airy-inf} .
\end{align}

\subsection{Energy current}
\label{app:en-curr-edge-air}

In order to study the behaviour of the energy current $J_E(x,t)$ near the edge $x\simeq v_{max} t$, the procedure is completely analogous to the one presented above; in particular, starting from Eq.~\eqref{operator2}, 
with $I(k,k')$ given by Eq.~(\ref{integralapp}) and $\varphi_{x,t}(k,k') = \phi_{x,t}(k) - \phi_{x,t}(k')$, we can expand (as done in Appendix~\ref{app:corr-edge}, see Eq.~\eqref{eq:app-exp-phi}) the phases $\phi_{x,t}(k)$ and $\phi_{x,t}(k')$  around the stationary point $k_s$, as we are interested in the leading-order correction to the space-time scaling limit.
Following the same steps as above, 
one finds Eq.~(\ref{eisler}), where $K^A$ is given by Eq.~\eqref{eq:KAXX-def}, 
obtained by taking the  limit $X \rightarrow Y$ of Eq.~\eqref{AiryKernel}.

As outlined in the main text, on the other hand, the expression in Eq.~(\ref{eisler}) vanishes when $h$ is set to its critical value $h_c=1$. Accordingly, one has to expand Eq.~(\ref{operator2}) up to the first non-vanishing order. In particular, for $h=h_c$,  the stationary point $k_s$ of the phase $\phi_{x,t}(k)$  
turns out to be $k_s = 0$, approached from above for $x\simeq v_{max} t$ and from below for   $x\simeq -v_{max} t$ (with $v_{max}=J$). Correspondingly, the odd derivatives of the dispersion relation $\varepsilon(k)$ become discontinuous at $k=0$ because $\varepsilon(k) = 2 v_{max} |\sin(k/2)|$, and therefore one has to consider the proper limits, i.e., 
\be 
\lim_{k \rightarrow 0^\pm} \varepsilon'(k) = v_g(0^\pm) = \pm v_{max} 
\quad \mbox{and} \quad \lim_{k \rightarrow 0^\pm} \varepsilon^{(3)}(k) = \mp v_{max}/4,
\label{thirdderivative}
\ee
while all the even derivatives vanish. 
As a consequence,
by expanding up to third order in the phase $\phi_{x,t}(k)$, for $x\simeq v_{max}t$, one finds (instead of Eq.~\eqref{eq:app-exp-phi} with $k_s = 0$)
\be
\begin{split}
\phi_{x,t}(k) & = k(v_{max} t-x) - \frac{1}{3!}\frac{v_{max}}{4} k^3 t + {\cal O}(k^{5}) \\ 
              & = -\frac{1}{3}K^3 -XK + {\cal O}\left(t^{-2/3}\right),
\end{split}
\ee
where we defined $k= (v_{max}t/8)^{-1/3} K$ and the scaling variable $X=(x - v_{max}t)/(v_{max}t/8)^{1/3}$ which is analogous to Eq.~(\ref{scalingvariables}), except for a numerical factor due to the fact that  $\varepsilon^{(3)}(k_s)$
at the critical point  is no longer $-v_{max}$ as in the non-critical case, but it is given by Eq.~(\ref{thirdderivative}). 
Analogous expansion is done for $\phi_{x,t}(k') =  -Q^3/3 -XQ + {\cal O}(t^{-2/3})$, where $k' = (v_{max}t/8)^{-1/3} Q$. 
Keeping into account that the factor $e^{-ik'}-e^{ik}$  in Eq.~(\ref{operator2}) must be expanded up to first order in $K$ and $Q$, since it vanishes identically at the lowest order, 
\be
e^{-ik'}-e^{ik} = -i(K+Q)(v_{max}t/8)^{-1/3} + {\cal O}\left(t^{-2/3}\right),
\ee 
we get the following expression for Eq.~(\ref{operator2})
\be
\begin{split}
J_E (X,t) =& \left(\frac{8}{v_{max}t}\right)^{2/3} \frac{J^2}{4} \int_{-\infty}^{+\infty} \frac{dK}{2 \pi} \int_{-\infty}^{+\infty} \frac{dQ}{2 \pi}  (Q+K) \; e^{-iKX-iK^3/3} e^{iQX+iQ^3/3} \\ 
&\times\left[\frac{f_{\beta_l}(K(v_{max}t/8)^{-1/3})+ f_{\beta_l}(Q(v_{max}t/8)^{-1/3})}{i(K-Q-2i\delta)}- \frac{f_{\beta_r}(K(v_{max}t/8)^{-1/3})+ f_{\beta_r}(Q(v_{max}t/8)^{-1/3})}{i(K-Q-2i\delta)} \right].
\end{split}
\label{eq:app-K-2}
\ee
In order to determine the first non-vanishing order, one needs to expand the Fermi-Dirac distributions
\be
\begin{split}
f_{\beta_l}(K(v_{max}t/8)^{-1/3})+ f_{\beta_l}(Q(v_{max}t/8)^{-1/3})-f_{\beta_r}(K(v_{max}t/8)^{-1/3})- f_{\beta_r}(Q(v_{max}t/8)^{-1/3})) &= \\
= [f_{\beta_l}'(0^+)-f_{\beta_r}'(0^+)](K+Q)(v_{max}t/8)^{-1/3} + {\cal O}\left(t^{-2/3}\right)
\end{split}
\label{eq:app-K-3}
\ee
with $f_\beta (0^+)= -\beta v_{max}/4$. 
By combining Eqs.~\eqref{eq:app-K-2} and \eqref{eq:app-K-3}, one eventually finds 
\be 
J_E(X,t) = \frac{1}{t} \frac{v_{max}^2 (\beta_r - \beta_l)}{2} K^c(X),
\ee 
where
\be
K^c(X) = \int_{-\infty}^{+\infty} \frac{dK}{2 \pi} \int_{-\infty}^{+\infty} \frac{dQ}{2 \pi}  \frac{(Q+K)^2 \;  e^{-iKX-iK^3/3} e^{iQX+iQ^3/3}}{i(K-Q-2i\delta)}.
\ee
One can make the expression of $K^c$ more explicit by taking a derivative with respect to $X$ and then by using Eqs.~\eqref{eq:def-Airy}, \eqref{eq:diff-eq-Airy}, and \eqref{eq:KAXX-def} in order to show that
\be
\frac{\partial K^c(X)}{\partial X} = -2 K^A (X,X),
\ee
which renders Eqs.~(\ref{eq:rel-kernels}) and, by integration, Eq.~\eqref{kernelcritical}.

\end{appendices}

\twocolumngrid

\end{document}